\input epsf.tex

\rightline{DAMTP-2009-62}
\vskip 30pt
\centerline{\bf TRIALITIES AND EXCEPTIONAL LIE ALGEBRAS:}
\vskip 3pt
\centerline{\bf DECONSTRUCTING THE MAGIC SQUARE}
\vskip 20pt

\centerline{Jonathan M.~Evans}
\vskip 5pt
\centerline{\it DAMTP, University of Cambridge}
\centerline{\it Centre for Mathematical Sciences}
\centerline{\it Wilberforce Road, Cambridge CB3 0WA, U.K.}
\vskip 5pt
\centerline{\tt J.M.Evans@damtp.cam.ac.uk}

\vskip 40pt

\font\bbb=msbm10
\font\abs=cmr10 at 9pt
\def\R{{\hbox{\bbb R}}}
\def\C{{\hbox{\bbb C}}}
\def\H{{\hbox{\bbb H}}}
\def\O{{\hbox{\bbb O}}}
\def\K{{\hbox{\bbb K}}}
\def\Z{{\hbox{\bbb Z}}}

\def\half{{\textstyle{1 \over 2}}}
\def\quarter{{\textstyle{1 \over 4}}}

\def\ndg{\hbox{\kern.05em}}
\def\mdg{\hbox{\kern-.05em}}
\def\wedgesq{\wedge^{\mdg 2 \ndg}}
\def\V{ {\rm V}}
\def\S{ {\rm S}}
\def\T{ {\rm T}} 
\def\U{{\rm U}}
\def\B{{\rm B}}
\def\g{{\bf g}}
\def\p{{\bf p}}
\def\h{{\bf h}}

\def\v{{\scriptscriptstyle {\rm V}}}
\def\e{e_\v}
\def\eps{\varepsilon}

\def\r{\sigma}
\def\a{\alpha}
\def\b{\beta}
\def\c{\gamma}
\def\d{\delta}
\def\ad{{\dot \alpha}}
\def\bd{{\dot \beta}}
\def\cd{{\dot \gamma}}
\def\dd{{\dot \delta}}
\def\vpb{\vphantom{\dot \beta}} 
 
\def\lp{{\! \! +}}
\def\lm{{\! \! -}}
\def\sp{ {\scriptscriptstyle +} }
\def\sm{ {\scriptscriptstyle -} }
\def\spm{ {\scriptscriptstyle \pm} }
\def\smp{ {\scriptscriptstyle \mp} }

\noindent 
{\abs A construction of the magic square, and hence of 
exceptional Lie algebras, is carried out using trialities 
rather than division algebras. By way of preparation,
a comprehensive discussion of trialities is given, 
incorporating a number of novel results and proofs.
Many of the techniques are closely related to, or derived from, 
ideas which are commonplace in theoretical physics. 
The importance of symmetric spaces in the magic 
square construction is clarified, allowing 
the Jacobi property to be verified for each algebra in 
a uniform and transparent way, with no detailed calculations 
required. A variation on the construction, 
corresponding to other symmetric spaces, is also given.}
\vskip 40pt 

\openup 0.5\jot

\noindent
{\bf 1. Introduction}
\vskip 20pt

\noindent
An understanding of the construction and classification of 
Lie groups and Lie algebras is essential in 
many branches of theoretical or mathematical
physics, and certainly in gauge theory, 
string theory, supersymmetry and integrable systems. 
Valuable insights into the Cartan-Killing classification 
[1,2] can be gained by approaching it  
via the normed division algebras of real numbers
$\R$, complex numbers $\C$, quaternions $\H$ and octonions $\O$
(and their classification in turn by Hurwitz's Theorem)---see 
e.g.~[3] and references therein.
The occurrence of the $so$, $su$ and $sp$ families of classical 
Lie algebras can then 
be attributed directly to the existence of $\R$, $\C$ and $\H$,
respectively, and it is natural to suppose that the five 
simple exceptional algebras should arise from $\O$ in some fashion. 
But while $G_2$ duly emerges as the automorphism group of the
octonions, the remaining exceptional algebras pose more of a  problem: 
how, exactly, should these things be understood in terms of octonions, 
or in terms of some related `exceptional' structure?

The magic square provides an answer to this question. It has 
appeared in various incarnations over the years, and these 
have been reconciled only comparatively recently, with the work of 
Barton and Sudbery [4]. 
The common starting point in any version of the magic square 
is a pair of division algebras (which are used to label
the rows and columns of the square) from which a Lie algebra can be 
constructed, with the following results:
$$
\matrix{
   &  \R    &  \C    &  \H    &  \ \ \ \ \ \O  \cr
\R \ \ \ \ \ \ &  so(3) &  su(3) &  sp(3) &  \ \ \ \ \ F_4 \cr
\C \ \ \ \ \ \ &  su(3) &  \! \! su(3) \oplus su(3) \! \! & su(6)
& \ \ \ \ \ E_6 \cr
\H \ \ \ \ \ \ &  sp(3) &  su(6) &  so (12) &  \ \ \ \ \ E_7 \cr
\O \ \ \ \ \ \ &  F_4   &  E_6   &  E_7   &  \ \ \ \ \ E_8 \cr
}
\eqno (1.1) $$
Some versions of the construction are not obviously symmetrical in 
the two division algebras and then part of the magic of the square
is its symmetry. For references to the earlier literature, see [4,3].

The principal aim of this paper is to give a description of the 
magic square, and hence of exceptional Lie algebras, in terms 
of {\it trialities} rather than division algebras. A triality is an 
algebraic structure---actually a multilinear map on three vector spaces 
to the reals---with an associated notion of symmetry [1]. There 
is a one-to-one correspondence between (normed) trialities and (normed)
division algebras and, in particular, the octonions are associated
with the celebrated symmetry properties of $so(8)$ [1,3].
One can therefore regard trialities and division 
algebras as being manifestations of a common, exceptional kind 
of mathematical structure. 

The symmetry of a triality---the triality group or algebra---is
central to the magic square, 
as is emphasised in the Barton-Sudbery approach [4].
This group is present whether one works with the triality as an 
algebraic concept in its own right, or whether one chooses to 
express things using division algebras. 
But the passage from a triality to a division algebra involves 
relinquishing some manifest symmetry, at least to the extent 
that a particular element of each vector space must be 
selected to play the role of the identity (the first 
step in defining a multiplication on the vector 
space and turning it into a copy of the division algebra) [1]. 
For this reason alone it is natural to ask whether there is something 
to be gained from carrying out a construction of 
the magic square using just trialities and avoiding 
division algebras explicitly. 
There is no doubt this should be possible,
but the question is whether it is illuminating. 
We shall endeavour to show that it is.

Using trialities instead of division algebras also seems 
particularly natural from the standpoint of mathematical physics, since 
trialities can be thought of as rather special examples of 
Dirac gamma matrices or invariant tensors [1,3,5]. 
Their distinguishing feature is an enhanced symmetry between 
vectors and spinors in the Euclidean dimensions concerned, 
namely $n= 1$, 2, 4, or 8.
This is interesting in its own right, but it 
also provides a link between the classification of trialities, 
or division algebras, and super-Poincare symmetry in dimensions 
3, 4, 6 and 10 [5,6,7] (see e.g.~[8,9,10] for background on supersymmetric
systems). The early papers in this area considered both 
super Yang-Mills [5,6,7] and classical superstrings or extended
objects [5,11,12,13] but interesting links between supersymmetry and 
exceptional structures of various kinds continue to be 
explored (see [10] and also [14,15,16] for a few more recent examples).

We shall give a thorough account of trialities 
and their symmetries in sections 3, 4 and 5 below.
Many of these results are well-known in some form, 
but we will provide a number of new derivations, 
including a simple proof of what is
sometimes called the {\it Principle of Triality} [1,3,4,17]. 
Our intention will be to make 
the connections with spinors as clear as possible, and although 
our treatment will be mathematically self-contained, many of the 
key ideas should be familiar in the context 
of theoretical physics.

Once we have laid the necessary foundations by investigating 
trialities, the construction of the magic square can be 
carried out remarkably simply, as we will see in section 6. 
Our work will also fully 
reconcile the magic square with other, superficially different 
ways of building Lie algebras which can be found in
the literature.
The techniques in question have been developed and 
applied in typically powerful fashion in the notes of
Adams [1], and similar ideas have been popularised for physicists in 
e.g.~Green, Schwarz and Witten [8]. These ideas 
really amount to a general method of approach 
rather than a uniform construction, however.
We will review this in some generality in section 2,
but we will also outline some aspects of it now, 
so that we can better explain the role it will play later 
in the paper.

Consider a Lie algebra $\h$ and a representation of it on a vector
space $\p$. We can attempt to define a new Lie algebra with vector space 
$\g = \h \oplus \p$ by first extending the Lie 
bracket, using the representation, and then checking the Jacobi
identity, with all definitions covariant under $\h$. 
Success is not automatic, and depends on certain properties of the
representation chosen, but there are plenty of examples where
it works. We can always extend $so(n)$ by its vector
representation to obtain $so(n{+}1)$, for instance.
This is an example of a symmetric pair,
$\g \supset \h$, meaning that the quotient of the 
corresponding groups $G/H$ is a symmetric space [2].
Any symmetric pair can be constructed in this way, 
provided we know the right representation to choose. 
The same general approach can be pursued even
if $\g \supset \h$ is not a symmetric pair, but this leaves 
much greater freedom in the definition of the bracket,
and verification of Jacobi is usually 
significantly more involved.

The symmetric space approach works nicely for certain exceptional 
algebras: we can build  $\g = F_4$ or $\g = E_8$ by starting from 
$\h = so(9)$ or $\h = so(16)$ and adding 
$\p$ in a suitable spinor representation, for example.
In either case, the Jacobi identity can be checked by using 
a Fierz rearrangement involving the $so(9)$ or $so(16)$ gamma
matrices [8]---a standard technique in QFT---or by various other
means [1]. The calculations are not too difficult, but they do not
provide much insight into why things work for these particular choices
of $\h$ and $\p$ (as remarked upon in e.g.~[3]). 
In contrast to the `generic' behaviour of our 
previous example, in which $\g = so(n{+}1)$ is constructed from 
$\h = so(n)$ for any $n$, there is something `exceptional' in these 
new examples, as befits the exceptional groups. 

A natural suspicion is that the triality 
corresponding to $so(8)$ somehow lies behind the success 
of the $F_4$ and $E_8$ symmetric space constructions,
and we can investigate this by looking at the simpler of these 
examples in more detail. If we try to build 
$F_4$ directly from $\h = so(8)$ we need to add 
$\p = \V\oplus \S_+ \oplus \S_-$, where $\V$, $\S_+$, $\S_-$ denote
vector and spinor representations, which are 
related by triality symmetries. The apparent disadvantage
of this approach is that $\g \supset \h$ is not a symmetric pair, the  
representation $\p$ is reducible, and so completing the definition of
the bracket and then checking Jacobi would seem to involve 
considerable further effort. By contrast, the symmetric pair 
approach amounts to re-grouping summands by taking 
$\h = so(8) \oplus \V = so(9)$
and $\p = \S_+ \oplus \S_-$, which becomes irreducible as a representation of 
this larger algebra. The definition of the bracket is then fixed,
but in re-grouping we have relinquished some symmetry by treating $\V$
differently from $\S_\pm$, and so the significance of the 
triality in verifying the Jacobi property is obscured.

The key to a more satisfactory understanding is the realisation 
that we can combine the best of both points of view, as follows.
We know that certain simplifications occur in building 
$F_4$ from $so(8)$ if we first extend the subalgebra to $so(9)$;
we should also note that this is an example of the `generic' sort 
of extension discussed above for $so(n)$ algebras---there is nothing 
very special about this in itself.
But the triality symmetry of $so(8)$ implies that we can
carry out such an extension in three inequivalent ways, 
by adding in $\V$, $\S_+$ or $\S_-$, and this is certainly 
special.
Despite the fact that for $\g = F_4$ and 
$\h = so(8)$ we are not dealing with a symmetric pair, 
the existence of these three intermediate $so(9)$ 
extensions fixes much of the Lie bracket on 
$\p = \V\oplus \S_+ \oplus \S_-$, as well 
as ensuring that large parts of the Jacobi identity 
are automatic. It turns out that the triality structure also 
provides an entirely natural way of completing the definitions
of the brackets, and we need do little more than write these down 
to check the remaining parts of the Jacobi identity.

To re-iterate: we can build 
$$
F_4 \supset so(9) \supset so(8)
\eqno(1.2) 
$$
by considering {\it three} intermediate algebras, related by triality
symmetry, with each inclusion corresponding to a symmetric pair.
Since there are essentially no calculations required in checking the 
Jacobi property (no Fierz-style computations, for instance) 
the success of the construction is as transparent as possible.
We can also see clearly how its `generic' aspects, 
involving orthogonal algebras, have been combined with 
the more `exceptional' properties of $so(8)$.
We will describe in section 6 how a similar approach   
works for each triality and produces the first row 
or column of the magic square. We will then explain how to
generalise it to a {\it pair} of trialities, leading to the 
complete magic square.
The full construction will be entirely uniform, 
treating all entries of the square on the same footing, 
with no special case-by-case analysis required.  
It will also share the desirable feature of 
the example above by allowing a clear separation of 
`generic' and `exceptional' features, so that we can see at which 
points the trialities are really put to use.

The paper concludes, in section 7, with a 
short summary followed by a few additional developments.
The first of these offers a slightly different perspective 
on the magic square, simplifying some aspects of it yet further.
Next we give a sketch of the relationship between
trialities and division algebras.
This enables us to discuss, in passing, how $G_2$ arises in the 
context of trialities---thus completing the set of 
exceptional algebras---but our main concern is to link 
the construction in this paper to the work in [4].
The last topic is a variation on the 
magic square: from one point of view, it corresponds 
to applying similar ideas to different symmetric pairs,
but it also has an attractive interpretation as a 
`folding' of the entries on the diagonal of the square.
This leads to an alternative, triality-based 
description of $E_7$. 
\vskip 50pt

\noindent
{\bf 2. General and Generic Constructions}
\vskip 20pt

\noindent
Throughout this paper we will confine our attention to real 
Lie algebras equipped with an invariant, positive-definite 
inner-product.
Every such algebra is a direct sum of simple algebras and a centre,
with the inner-product restricting to some multiple of the Killing 
form on each simple factor [2,18]. 
Lie algebra elements in the abstract will be denoted by $T$ or
$X$, whereas $M$ will be reserved for concrete representations in
terms of matrices. We shall frequently work with component or 
index notation corresponding to orthonormal bases, and the summation
convention will apply to any repeated index. In keeping with our use 
of terminology from theoretical physics, we will also 
frequently refer to basis elements of a Lie algebra as {\it generators\/}.
Other standard notation includes e.g.~$\oplus$ to denote the
orthogonal direct sum of vector spaces (but not necessarily 
a direct sum of algebras) and 
$\wedge^p$ to denote the antisymmetrised tensor product of rank $p$.
\vskip 20pt

\noindent
{\bf 2.1 General constructions: cosets and symmetric pairs} 
\vskip 10pt

\noindent
Let $\h$ be a real Lie algebra with orthonormal 
basis $\{ T_A \}$ and Lie bracket 
$$
\wedgesq \h \, \rightarrow \, \h \ , 
\qquad \qquad [ \, T_A , \, T_B \, ] \, = \, f_{ABC} \, T_C 
\eqno(2.1) $$
The structure constants $f_{ABC}$ are real numbers,
antisymmetric in the indices $ABC$, which is to say that 
the bracket and inner-product (denoted by round brackets) 
combine to give a map 
$$
f \, :  \, \wedge^{\mdg 3 \ndg} \h \, \rightarrow \, \R \ ,
\qquad \quad f_{ABC} \, = \, ( \, T_A , \, [T_B , T_C] \, ) 
\eqno (2.2) $$
The Jacobi identity for $\h$ is a condition on double brackets of 
elements of the form $TTT$, namely 
$$
[ \, T_A , \, [ T_B , T_C ] \, ]  \, + \,  
[ \, T_B , \, [ T_C , T_A ] \, ]  \, + \, 
[ \, T_C , \, [ T_A , T_B ] \, ]  \, = \, 0 \ 
\eqno(2.3{\rm a}) $$
which is equivalent to 
$$
f_{BCD}^{\phantom{i}} \, f_{ADE}^{\phantom{i}} \, + \, 
f_{CAD}^{\phantom{i}} \, f_{BDE}^{\phantom{i}} \, + \,  
f_{ABD}^{\phantom{i}} \, f_{CDE}^{\phantom{i}} \, = \, 0 
\eqno(2.3{\rm b}) $$
This is the adjoint representation of the algebra,
expressed in components; it is also the 
statement that the map in (2.2) is invariant
under the adjoint action.

Let $\p$ be a real vector space with a positive-definite inner-product
and $\{ X_i \}$ an orthonormal basis. Suppose that 
there is a representation of $\h$ on this space 
by real matrices $(M_A)_{ij}$ antisymmetric in $ij$, 
implying that the inner-product is invariant.
This provides us with an extension of the 
Lie bracket because we can define 
$$ 
\h \otimes \p \, \rightarrow \, \p \ , 
\qquad \qquad
[ \, T_A , \, X_i \, ] \, = \, - (M_A)_{ij} X_j 
\eqno(2.4)
$$ 
The fact that we have a representation
$$
(M_A)_{ij} \ndg (M_B)_{jk} \, - \, (M_B)_{ij} \ndg (M_A)_{jk} \ = \ f_{ABC} \ndg
(M_C)_{ik}
\eqno(2.5{\rm a})$$
is then equivalent to the Jacobi property for 
elements of the form $TTX$, namely
$$
[ \, T_A , \, [ T_B , X_i \, ] \, ]  \, - \, 
[ \, T_B , \, [ T_A , X_i \, ] \, ]  
\, = \,  
[ \, [ T_A , T_B ] , \, X_i \, ]  
\eqno(2.5{\rm b}) $$
 
Our aim is to understand the conditions under which  
$\g  = \h \oplus \p $ 
can be made into a Lie algebra. To do this we must 
complete the definition of the Lie bracket between elements in $\p$
and then check the Jacobi property for the remaining 
combinations of generators $TXX$ and $XXX$.
The representation of $\h$ on $\p$ may be reducible, but 
it must decompose as a sum of real irreducible pieces. 
Using the freedom to scale the basis elements and inner-products 
on each irreducible representation, we can ensure 
that $\{ T_A , X_i \}$ is orthonormal with respect to 
an invariant inner-product on $\g$. Thus, without loss of generality,
$$
( T_A , T_B ) = \delta_{AB} \ , \qquad 
( T_A , X_i ) = 0 \ , \qquad 
( X_i , X_j ) = \delta_{ij}
\eqno(2.6) $$

It then follows that the structure constants of the algebra we are seeking
to construct are totally antisymmetric, and (2.4) implies 
$$
( \, T_A , \, [\, X_i , \, X_j \, ] \, )
\, = \, ( \, X_j , \, [\, T_A , \, X_i \, ] \, )
\, = \, - (M_A)_{ij} 
\eqno (2.7) $$
The most general form for the remaining part of the Lie bracket is 
therefore
$$
\wedgesq {\bf p} \, \rightarrow \, \g  \ , \qquad \qquad  
[ \, X_i , \, X_j \, ] \, = \, - (M_A)_{ij} T_A  \, + \, c_{ijk} X_k
\eqno(2.8)$$ 
Here, the representation matrices $(M_A)_{ij}$ are 
being re-interpreted as a map $\wedgesq \p \rightarrow \h$ 
(by dualising spaces using the inner-products) 
and this is covariant under $\h$ by virtue of (2.5). 
The second term in the bracket sends 
$\wedgesq \p \rightarrow \p$, but since 
$c_{ijk}$ must be antisymmetric in $ijk$ 
it actually define a map 
$$
c \, : \, \wedge^{\mdg 3 \ndg} \p \, \rightarrow \, \R \ ,
\qquad \quad  c_{ijk} \, = \, ( \, X_i , \, [X_j , X_k] \, ) 
\eqno (2.9) $$
It is easy to check that the condition for this map to 
be $\h$-invariant 
$$
(M_A)_{i \ell} \, c_{\ell j k}
\, + \, (M_A)_{j \ell} \, c_{i \ell k}
\, + \, (M_A)_{k \ell} \, c_{i j \ell} \, = \, 0 
\eqno (2.10{\rm a}) $$
is necessary and sufficient for the Jacobi property to hold for 
combinations of elements $TXX$, or
$$
[ \, T_A , [X_i , X_j ] \, ] 
\, = \, [ \, [T_A, X_i ] , \, X_j \, ] \, + \, 
[\, X_i , \, [ T_A , X_j ] \,]
\eqno(2.10{\rm b}) $$

The final step is to check the Jacobi property for elements of type 
$XXX$, and the definition of $\g$ will be complete iff
$$
[ \, X_i , \, [ X_j , X_k ] \, ]  \, + \,  
[ \, X_j , \, [ X_k , X_i ] \, ]  \, + \, 
[ \, X_k , \, [ X_i , X_j ] \, ]  \, = \, 0 
\eqno(2.11) $$
This amounts to a non-trivial quadratic 
relationship between $(M_A)_{ij}$ and $c_{ijk}$, in general. 
The situation becomes significantly simpler in the special 
case $c_{ijk} = 0$, however, with (2.8) then reducing to 
$$
\wedgesq {\bf p} \, \rightarrow \, \h \ , \qquad \qquad 
[ \, X_i , \, X_j \, ] \, = \, - (M_A)_{ij} T_A  
\eqno(2.12)$$ 
In these circumstances the entire construction 
is determined by the choice of representation of $\h$ on $\p$,
and the final part of the Jacobi condition (2.11) holds iff 
$$
(M_A)_{ij \vpb} (M_A)_{k\ell \vpb} \, + \, (M_A)_{jk \vpb} (M_A)_{i\ell \vpb} 
\, + \, (M_A)_{ki \vpb} (M_A)_{j \ell\vpb}
\, = \, 0
\eqno(2.13)
$$
Note the $\Z_2$ grading corresponding to $\g = \h \oplus \p$ 
which is manifest in the form of the brackets (2.1), (2.4) and (2.12);
this is what is meant by $\g \supset \h$ being a symmetric pair [2].

Whether we are attempting to build a new algebra as part of a 
symmetric pair or by using the more elaborate possibility involving
a non-zero map $c$, the identification of $\g$ is 
usually fairly straightforward.
In particular, if $\h$ is a subalgebra of maximal rank,
then the roots of $\g$ are the roots of $\h$ together with 
the weights of $\p$. This approach 
can be used to build up each of the classical series of 
Lie algebras by adding in the defining or fundamental representations
of smaller algebras to make larger ones. 
There are some well-known families of symmetric pairs 
of orthogonal algebras which we will need 
to describe in detail, in preparation for the 
work to follow.
\vskip 20pt

\noindent
{\bf 2.2 The cases $so(n{+}1) \supset so(n)$ and $so(n{+}n') \supset 
so(n) \oplus so(n')$} 
\vskip 10pt

\noindent
Let $\V$ be a real vector space of dimension $n$ with a 
positive-definite inner-product (denoted by a dot). There is 
a natural action 
$$
\wedgesq \V \, : \, \V \rightarrow \V \ , \qquad 
(u\wedge w) (v) \, =  \, u (w \cdot v) - w (u \cdot v)  
\eqno(2.14) $$
which enables us to identify 
$$
so(n) = \wedgesq \V
\eqno(2.15) $$
with $\V$ the defining, or fundamental, or vector representation 
of $so(n)$. This is consistent with regarding the 
right-hand side as the adjoint representation, which 
can always be identified with the algebra itself. 
Now by taking $\V \oplus \R$ as the fundamental representation of
$so(n{+}1)$ we have, similarly,
$$
so(n{+}1) \, = \, \wedgesq ( \V \oplus \R) \, 
= \, \wedgesq \V \oplus \V \, = \, so(n) \oplus \V
\eqno (2.16)$$
The direct sum corresponds to a $\Z_2$ grading of the 
resulting Lie brackets and so $so(n{+}1) \supset so(n)$
is a symmetric pair. It will also be very useful for us to 
understand this in a slightly different way, however.

Following the procedure of the subsection above, consider 
$\h = so(n)$ and $\p = \V$ with representation matrices $(M_A)_{ij}$. 
The normalisation of the 
generators of $so(n)$ can be fixed by demanding 
$$
\half \ndg (M_A)_{ij} (M_B)_{ij} \, = \, \delta_{AB}
\eqno(2.17) $$
and this in turn fixes the multiple of the Killing form 
we are using, since the inner-product of basis elements is 
$(T_A , T_B ) = \delta_{AB}$, by assumption.
Now any antisymmetric $n{\times}n$ matrix can be written
$K_{ij} = K_A (M_A)_{ij}$ and, because (2.17) then determines the 
coefficients to be $K_A = {1 \over 2} (M_A)_{k\ell} K_{k\ell}$, 
we must have
$$
(M_A)_{ij} (M_A)_{k \ell} 
\, = \, 
\delta_{ik} \ndg \delta_{j \ell} - \delta_{i \ell} \ndg \delta_{jk}
\eqno(2.18) $$
This is enough to ensure that the symmetric pair
construction works, since the key condition (2.13) follows
immediately. Thus, $\g = \h \oplus \p$ is a Lie algebra with
brackets defined by (2.4) and (2.12) and on closer inspection 
we find $\g = so(n{+}1)$ once again (see also below).

To understand how these two lines of argument are related, 
recall that the representation matrices $(M_A)_{ij}$ 
define invariant maps: $\wedgesq \V \rightarrow so(n)$ and its
transpose $so(n) \rightarrow \wedgesq \V$ (as pointed out after 
(2.8)). The conditions (2.17) and (2.18) say precisely that these 
maps are mutually inverse (and hence orthogonal) and so together 
they implement the identification (2.15).
In scrutinising the details, we note 
that when an antisymmetric pair of indices $ij$ is used as a basis, 
a sum over all values of $i$ and $j$ amounts to summing over the basis twice. 
The coefficient of $\half$ included in (2.17) is therefore 
just what we need to compensate for the double counting implicit in
such a sum. Furthermore, the expression on the right 
of (2.18) is exactly the identity map on $\wedgesq \V$ when a similar 
convention is adopted in writing down its action on this space.
 
The identification (2.15) also lies behind another 
standard description of $so(n)$ in which basis elements 
are labelled by antisymmetric pairs of vector indices:
$$
T_{ij} = (M_{A})_{ij} T_A 
\eqno(2.19) $$
With this definition, a short calculation 
(using (2.1), (2.5) and (2.18)) yields the algebra in familiar form 
$$
[ \, T_{ij} , \, T_{k\ell} \, ]  \, = \, 
- \delta_{ik} T_{j\ell}  + \delta_{jk} T_{i\ell}
- \delta_{j\ell} T_{ik} + \delta_{i\ell} T_{jk}
\eqno(2.20) $$
and on taking matrix representations of both generators $T$
in (2.19) we deduce 
$$\eqalign{ 
(M_{ij})_{k \ell} \, & =  \, (M_A)_{ij} (M_A)_{k \ell} \cr
& = \, \delta_{ik} \delta_{j\ell} - \delta_{i\ell} \delta_{jk} }
\eqno(2.21)
$$
The new basis in (2.19) must be orthonormal, given our 
interpretation of (2.17) and (2.18) as orthogonality conditions, 
but this can be checked:
$$
(T_A , T_B ) = \delta_{AB} \qquad \Rightarrow \qquad 
( T_{ij} , T_{k \ell} ) \, = \, \delta_{ik} \delta_{j \ell} 
- \delta_{i \ell} \delta_{jk}
\eqno(2.22) $$
It is also instructive to re-write the brackets 
(2.4) and (2.12) as 
$$
[\, T_{ij} , \, X_k \, ] \, = \, - \delta_{ik} X_j \, + \,
\delta_{kj} X_i \ , \qquad 
[\, X_i , \, X_j \, ] \, =  \, - T_{ij} 
\eqno(2.23)$$
We need only take the additional generators 
to be $T_{n{+}1 \, i} = X_i$ in order to 
cast $\g = so(n{+}1)$ in the same standard form 
(2.20) as the original algebra $\h = so(n)$.

It is straightforward to extend the discussions above 
to symmetric pairs 
$so(n{+}n') \supset so(n) \oplus so(n')$.
If $\V$ and $\V'$ are the fundamental representations  
of $so(n)$ and $so(n')$ then 
$$
so(n) = \wedgesq  \V \ , \qquad 
so(n') = \wedgesq \V' 
\eqno (2.24) 
$$
according to (2.15). But we can also regard $\V \oplus \V'$ 
as the fundamental representation of $so(n{+}n')$ so that,
similarly, 
$$
so(n{+}n') \, = \, \wedgesq (\V \oplus \V') 
\eqno (2.25) $$
Then since 
$$
\wedgesq (\V \oplus \V') 
\, = \, \wedgesq \V \oplus \wedgesq \V' \oplus (\V \otimes \V')
\eqno (2.26) 
$$
we deduce that 
$$
so(n{+}n') \, = \, so(n) \oplus so(n') \oplus (\V \otimes \V')
\eqno (2.27) 
$$
and the Lie brackets are readily seen to be consistent with a $\Z_2$ 
grading. 

Once again, it will be very helpful for what follows later in the
paper to see how the same conclusion is 
reached by applying the general approach of the last subsection.
We will modify our notation slightly, to reflect the fact that 
we are now starting with a direct sum, writing 
$$
\h = so(n) \oplus so(n')
\quad 
{\rm with~basis} 
\quad 
\{ \, T_A , \, T_{A'} \, \} 
\eqno (2.28) $$
If $\V$ and $\V'$ have bases labelled by $i$ and $i'$ then we take
$$
\p = \V \otimes \V'
\quad 
{\rm with~basis} 
\quad 
\{ X_{i \ndg i'} \} 
\eqno (2.29) $$
All bases are orthonormal and the 
representation matrices $(M_A)_{ij}$ and $(M_{A'})_{i' j'}$ for 
$so(n)$ and $so(n')$ will be chosen to satisfy 
$$
\half \, (M_A)_{ij} (M_B)_{ij} \, = \, \delta_{AB}
\ , \qquad 
\half \, (M_{A'})_{i'j'} (M_{B'})_{i'j'} \, = \, \delta_{A'B'}
\eqno(2.30) $$
It follows, as before, that they also obey
$$
(M_A)_{ij} (M_A)_{k \ell} 
\, = \, 
\delta_{ik} \delta_{j \ell} - \delta_{i \ell} \delta_{jk}
\ , \qquad 
(M_{A'})_{i'j'} (M_{A'})_{k' \ell'} \ 
\, = \, 
\delta_{i'k'} \delta_{j' \ell'} - \delta_{i' \ell'} \delta_{j'k'}
\eqno (2.31) $$

Given our slight change in notation, the definition (2.4) now
takes the form
$$
[ \, T_{A} , \, X_{i \ndg i'} \, ] \, = \,  - (M_{A})_{ij} X_{j i'} \ , 
\qquad 
[ \, T_{A'} , \, X_{i \ndg i'} \, ] \, 
= \, - (M_{A'})_{i'j'} X_{i \, j'} 
\eqno(2.32{\rm a}) $$
If we convert to bases of type (2.19) for both $so(n)$ and $so(n')$,
then Lie algebra indices $A$ and $A'$ are replaced 
by antisymmetric pairs $k\ell$ and $k'\ell'$ and we get 
$$\eqalign{
[ \, T_{k\ell} , \, X_{i \ndg i'} \, ] \, 
= \, - (M_{k\ell})_{ij} X_{ji'} 
\ & = \ - \, \delta_{ki} X_{\ell i'} + \delta_{\ell i} X_{ki'} 
\cr
[ \, T_{k' \ell'} , \, X_{i \ndg i'} \, ] \, 
= \, - (M_{k'\ell'})_{i'j'} X_{ij'} 
\ & = \ - \, \delta_{k'i'} X_{i \ell'} + \delta_{\ell' i'} X_{ik'}  
\cr 
} \eqno (2.32{\rm b})$$
The definition of the bracket is completed as in (2.12),
$$\eqalign{
[\, X_{i \ndg i'} , \, X_{jj'} \, ] \ & = \,
- \, (M_A)_{ij} \, \delta_{i'j'} \ndg T_A 
\, - \, 
\delta_{ij} \, (M_{A'})_{i'j'} \ndg T_{A'}
\cr
& = \ - \, \half \ndg (M_{k \ell})_{ij} \, \delta_{i'j'} \, T_{k \ell} 
\, - \,  
\half \ndg \delta_{ij} \, (M_{k' \ell'})_{i'j'} \, T_{k' \ell'}
\cr 
& = \ - \, T_{ij} \, \delta_{i'j'} \, - \,  
\delta_{ij} \, T_{i'j'}
\cr 
} \eqno (2.33)$$
To verify the Jacobi property, it is sufficient to check 
generators of the form $XXX$ and for
the case at hand a typical term is 
$$\eqalign{
[ \, [X_{i \ndg i'} , X_{jj'} ] , \, X_{kk'} \, ] 
\ & = \ (M_A)_{ij} (M_A)_{k \ell}  
\, \delta_{i' j'} \, X_{\ell k'}
\, + \, (M_{A'})_{i'j'} (M_{A'})_{k'\ell'} 
\, \delta_{i j} \, X_{k \ell'}
\cr
& = \ \delta_{ik} \, \delta_{i' j'} \, X_{j k'}
\, - \, \delta_{jk} \, \delta_{i' j'} \, X_{i k'}
\, + \, \delta_{i'k'} \, \delta_{i j} \, X_{k j'}
\, - \, \delta_{j'k'} \, \delta_{i j} \, X_{k i'}
\cr} 
\eqno(2.34) $$
It is easy to see that cycling in the 
pairs of indices $i \ndg i'$, $jj'$ and $kk'$ and adding gives zero, 
as desired. 

We have shown, therefore, that there exists a Lie algebra 
$\g = \h \oplus \p$ with brackets defined by (2.32) and (2.33);
the result is precisely (2.27) once again, so $\g = so(n{+}n')$.
Because this works for all positive integers $n$ and $n'$,
we will refer to it as a `generic' type of construction.
We will see in section 6 how it can be combined with 
certain `exceptional' features which arise only 
when $n$, $n' = 1, 2, 4$ or $8$ and how this leads 
to the magic square. But first we must discuss trialities.
\vskip 50pt

\noindent
{\bf 3. Trialities and Spinors}
\vskip 20pt

\noindent
The term `triality' is commonly used to refer to the exceptional 
outer automorphism symmetries of $so(8)$ 
(exceptional by comparison with all other simple Lie algebras)
and the ensuing relationships amongst its representations [17].
An alternative way of capturing this idea was 
advocated by Adams [1],
who defined trialities as algebraic structures involving 
three $n$-dimensional vector spaces and then went on to establish a 
one-to-one correspondence with $n$-dimensional division algebras. 
Any triality, in this sense, has a symmetry algebra $tri(n)$ 
with just the same kinds of outer automorphisms as $so(8)$,
thereby setting this most famous example in context, 
as the head of a family.
Its special features then arise because $tri(8) = so(8)$, which also 
allows them to be related in a natural fashion to the corresponding 
division algebra of octonions $\O$.  
For lower values of $n$ one finds  
$tri(4) = so(4) \oplus su(2)$, corresponding to $\H$, and 
$tri (2) = so(2) \oplus so(2)$, corresponding to $\C$,
and in these cases the symmetries of $tri(n)$ depend on the 
fact that the algebra is no longer simply $so(n)$.
The only other case allowed is rather degenerate, with $tri(1)$ trivial, 
corresponding to $\R$. (We will see how to derive the properties of
each triality in section 5.)

The approach to trialities in this paper rests firmly on the 
ideas and results of [1], but there will be a few significant 
departures and developments in both content and presentation.
We start by giving a new definition of a triality, which is 
easily shown to be equivalent to the 
notion of {\it normed triality} introduced in [1] but which 
seems to offer some advantages (for our purposes, at least).
To avoid confusion, we should also point out that Adams 
reserves the unqualified term {\it triality} for a distinct, weaker 
concept in which the vector spaces are not equipped with
inner-products---we shall not consider such things here. 
Our treatment of trialities and their symmetries will be
self-contained and will occupy us for the next three sections of the paper.

Let $\V$, $\S_+$, $\S_-$ be real vector spaces, each with a
positive-definite inner-product (the choice of names will 
be explained by what follows) and consider a trilinear map
$$
\gamma \, : \, \V \times \S_+ \times \S_- \ \rightarrow \ \R
\eqno(3.1{\rm a}) $$
If we fix a vector in any one of these spaces, then $\gamma$ will provide us
with linear maps between the remaining two, using the 
non-degeneracy of their inner-products. Thus, for any elements 
$u_\v$, $u_\sp$, $u_\sm$ in $\V$, $\S_+$, $\S_-$, we have maps 
$$
\r (u_\v) \, : \, \S_- \rightarrow \S_+ \ , 
\qquad \quad
\r (u_\sp) \ : \ \V \rightarrow \S_- \ , 
\qquad \quad
\r (u_\sm) \, : \, \S_+ \rightarrow \V 
\eqno(3.1{\rm b})$$
which are defined by demanding that the inner-products 
(denoted by dots) satisfy
$$
u_\sp \! \cdot ( \, \r (u_{\v}) u_\sm ) \ = \ 
u_\sm \! \cdot ( \, \r (u_\sp) u_\v \, ) \ = \ 
u_{\v} \cdot ( \, \r (u_\sm ) u_\sp ) \ = \ 
\gamma (\, u_\v, \, u_\sp , \, u_\sm \, ) 
\eqno(3.1{\rm c}) $$
for all $u_\v$, $u_\sp$, $u_\sm$. 
But it is also natural to require that these maps respect the
inner-products in some appropriate sense, which prompts the following
definition.
\vskip 3pt

A {\bf triality} is a map $\gamma$ as in (3.1) such that 
for any $u_\v$, $u_\sp$, $u_\sm$ of unit length, the corresponding 
maps $\r (u_\v)$, $\r (u_\sp)$, $\r (u_\sm)$ 
are orthogonal, or length preserving.
\vskip 3pt

An immediate consequence is that $\V$ and $\S_\pm$ must all have the 
same dimension $n$, but the concept of a triality is
much more restrictive than this. Suppose 
$$
\gamma' \, : \, \V' \times \S'{}_\lp \times \S'{}_\lm \ \rightarrow \ \R
\eqno (3.2)$$
is also a triality, with each space of dimension $n'$.
We say that $\gamma$ and $\gamma'$ are isomorphic 
if there exist invertible linear maps 
$$
R_\v  : \V \rightarrow \V' \ , \qquad 
R_\sp  : \S_+ \rightarrow \S'{}_\lp \ , \qquad 
R_\sm  : \S_- \rightarrow \S'{}_\lm 
\eqno(3.3{\rm a}) $$
(implying $n = n'$) under which the inner-products 
are invariant, and for which 
$$
\gamma' ( \, R_\v u_\v , \, R_\sp u_\sp , \, R_\sm u_\sm )
\, = \, \gamma  ( \, u_\v , \, u_\sp , \, u_\sm)
\eqno(3.3{\rm b}) $$
(Note that the composite map on the left hand side is indeed a
triality, for any set of orthogonal maps $R$.) Armed with this notion
of equivalence, we have the following fundamental result.
\vskip 10pt

\noindent
{\bf Classification Theorem:}
Trialities exist iff $n$ = 1, 2, 4 or 8 and they are unique up to isomorphism.
\vskip 5pt

\noindent
Proof: see the proposition, lemma and
corollary below, and the appendix.
\vskip 10pt

The correspondence with division algebras means that 
the existence of just four trialities matches exactly
with the existence and uniqueness of $\R$, $\C$, $\H$ and $\O$.
In [1], the link with division algebras is also utilised 
by writing down triality maps explicitly
for $n = 1$, 2, 4 in terms of $\R$, $\C$, $\H$. But the
$n=8$ case is dealt with differently (at least to begin with) 
by applying general results on spin representations. 
One motivation for treating the $n=8$ 
triality in this fashion is that it 
allows one to circumvent some of the complications inherent in the 
non-associative multiplication of $\O$. 

In this paper we will
work in terms of trialities throughout, returning to division algebras 
only briefly, in section 7, 
to outline connections with previous results.
From this point of view, the classification 
of division algebras can be regarded as a corollary of 
the Classification Theorem for trialities,  
rather than the other way around. 
We shall give a uniform treatment of all four cases,
with spinors playing a central role. The manner in which trialities 
can be realised in terms of spin representations is explained in 
[1,3,5], but the relationship is even more intimate than these
accounts might suggest. 
We will show that spinors emerge {\it inevitably} 
from the concept of a triality, with $\gamma$ becoming 
an $so(n)$-invariant map.

Let $u$ be a vector belonging to any one of the spaces 
$\V$, $\S_+$, $\S_-$. The map $\sigma (u)$ between the other two 
spaces is orthogonal whenever $u$ is of unit length, by definition. 
But since $\gamma$ is multilinear this implies
$$
\r (u ) \, \r ( u)^{\mdg \T}  = \, | u |^2 \ndg 1 \, , \qquad 
\r (u )^{\mdg \T} \r ( u) \, = \, | u |^2 \ndg 1 
\ 
\eqno (3.4{\rm a}) $$
for $u$ with {\it any\/} squared length $| u |^2 = u \mdg \cdot \mdg u$
(the identity operator on each space being denoted by 1).
Equivalent forms are obtained by polarising: combining the 
relations above as applied to $u$, $w$ 
(any other vector in the same space) and $u + w$ gives 
$$\eqalign{ 
\r (u ) \, \r ( w )^{\mdg \T} + \, 
\r (w ) \, \r ( u )^{\mdg \T} \, & = \ 2 \ndg u \mdg \cdot \mdg w \, 1 \cr
\r (u )^{\mdg \T} \r ( w ) \, \ndg + \, \ndg 
\r (w )^{\mdg \T} \r ( u ) \ & = \ 2 \ndg u \mdg \cdot \mdg w \, 1 \cr
}
\eqno (3.4{\rm b}) $$
Thus, within our definition of a triality lies 
a structure which closely resembles a Clifford algebra. Rather than
building spin representations from this as it stands, however,
it will be convenient for us to re-express things using index or 
component language.

Consider orthonormal bases for $\V$, $\S_+$, $\S_-$
with labels of type $a$, $\a$, $\ad$ (each taking $n$ values) 
reserved for each of these spaces. 
Since the nature of the index then also identifies
the space to which a vector belongs, we will write the components 
of $u_\v$, $u_\sp$ $u_\sm$ simply as $u_a$, $u_\a$, $u_\ad$, omitting the
additional subscripts to avoid cluttering our expressions.
With respect to these bases, $\gamma$ has 
a set of real components $\gamma_{a \ndg \a \ndg \ad}$ and the
matrices for the maps in (3.1b) are
$$
\r (u_\v)_{\a \ad} \, = \, u_a \ndg \gamma_{a \ndg \a \ndg \ad}  \ , 
\quad \quad
\r (u_\sp)_{\ad a} \, = \, u_\a \ndg \gamma_{a \ndg \a \ndg \ad}  \ , 
\quad \quad
\r (u_\sm)_{a \a} \, = \, u_\ad \ndg \gamma_{a \ndg \a \ndg \ad}  
\eqno(3.5)
$$
Clearly, (3.4) holds for any $u$ and $w$, in any one of the spaces, 
iff
$$\eqalignno{
\gamma_{ a \ndg \a \ndg \ad \vpb} \ndg \gamma_{ \ndg b \ndg \b \ndg \ad \vpb} 
\ + \ 
\gamma_{ \ndg b \ndg \a \ndg \ad \vpb}  \ndg \gamma_{ a \ndg \b \ndg \ad \vpb} 
& \ = \  
2 \ndg \delta_{ab \vpb} \ndg \delta_{\a \b \vpb}
& (3.6{\rm a}) \cr
\gamma_{ a \ndg \a \ndg \ad \vpb} \ndg \gamma_{ \ndg b \ndg \a \ndg \bd} 
\ + \ 
\gamma_{ \ndg b \ndg \a \ndg \ad \vpb}  \ndg \gamma_{ a \ndg \a \ndg \bd} 
& \ = \  
2 \ndg \delta_{ab \vpb} \ndg \delta_{\ad \bd \vpb}
& (3.6{\rm b}) \cr
\gamma_{ a \ndg \a \ndg \ad \vpb } \ndg \gamma_{ a  \ndg \b \ndg \bd} 
\ + \ 
\gamma_{ a \ndg \b \ndg \ad \vpb}  \ndg \gamma_{ a \ndg \a \ndg \bd} 
& \ = \ 
2 \ndg \delta_{\a \b \vpb} \ndg \delta_{\ad \bd \vpb}
& (3.6{\rm c}) \cr
}
$$
Together, these conditions constitute the definition of a triality 
in component form.

There is also something very important to be learnt 
about the logical connections between the individual equations 
in (3.6), which can be uncovered as follows.
For any $u_\v$ in $\V$ with $u_\v \mdg \cdot \mdg u_\v = u_a u_a =1$ we 
have
$$
\r(u_\v) \, \r(u_\v)^\T = \, 1 \qquad {\rm or}
\qquad 
\r(u_\v)_{\a \ad} \, \r(u_\v)_{\b \ad} \, = \,
u_{\vpb a} \gamma_{a \ndg \a \ndg \ad \vpb} \, 
u_{b \vpb} \gamma_{\ndg b \ndg \b \ndg \ad \vpb} 
\, = \, \delta_{\a \b \vpb} 
\eqno(3.7{\rm a}) $$
which is true iff (3.6a) holds; but also 
$$
\r(u_\v)^\T \r(u_\v) \, = \, 1 \qquad {\rm or} \qquad 
\r(u_\v)_{\a \ad \vpb} \, \r(u_\v)_{\a \bd} \, = \, 
u_{a \vpb} \gamma_{a \ndg \a \ndg \ad \vpb} \, 
u_{b \vpb} \gamma_{\ndg b \ndg \a \ndg \bd} \, = \, \delta_{\ad \bd} 
\eqno(3.7{\rm b}) $$
which is true iff (3.6b) holds. 
Now (3.7a) is equivalent to (3.7b) simply
because the matrix for $\r(u_\v)$ is square.
This means that 
(3.6a) is equivalent to (3.6b) just by virtue of $\S_+$ and $\S_-$
having the same dimension.
We can obviously apply identical reasoning with the spaces permuted,
so $\sigma (u_\sp)$ is orthogonal  
for any unit vector $u_\sp$ 
iff (3.6a) and (3.6c) hold, and $\sigma(u_\sm)$ is orthogonal
for any unit vector $u_\sm$ iff (3.6b) and (3.6c) hold.
The full significance of these remarks will become clear shortly, 
in justifying the Classification Theorem.

Now to discuss spinors, using the component definition (3.6).
The group $SO(n)$ which preserves the inner-product on $\V$ 
has a Lie algebra which we will denote by 
$so(n)_\v = \wedgesq \V$. Its generators can be written 
$T_{cd}$, labelled by antisymmetric pairs of vector indices, 
with the fundamental representation 
$$
(M_{cd})_{ab} \, = \, \delta_{ac} \ndg \delta_{bd} \, - \, 
\delta_{ad} \ndg \delta_{bc}
\eqno(3.8) $$
all as in section 2.
But once $a$ and $b$ are interpreted as 
vector indices, the conditions (3.6a) and (3.6b) 
are recognisable 
as Dirac/Clifford-like relations which allow the construction of
spin representations of $so(n)_\v$ on $\S_\pm$ along 
familiar lines. To be specific, the real antisymmetric matrices 
$$\eqalign{
(M_{cd})_{\a \b \vpb} 
\ \equiv \ 
\half (\gamma_{cd})_{\a \b \vpb} 
\ & \equiv \  
{\textstyle {1 \over 4}} 
(\, \gamma_{ c \ndg \a \ndg \ad \vpb} \gamma_{ d \ndg \b \ndg \ad \vpb}
- \gamma_{ d \ndg \a \ndg \ad \vpb}  \gamma_{ c \ndg \b \ndg \ad \vpb} \, ) 
\cr
(M_{cd})_{\ad \bd} \ \equiv \ \half (\gamma_{cd})_{\ad \bd} 
\ & \equiv \  
{\textstyle {1 \over 4}} (\, \gamma_{ c \ndg \a \ndg \ad \vpb} 
\gamma_{ d \ndg \a \ndg \bd \vpb}
- \gamma_{ d \ndg \a \ndg \ad \vpb}  \gamma_{ c \ndg \a \ndg \bd \vpb} \, ) 
\cr 
} \eqno (3.9) $$
each obey the same algebra as $(M_{cd})_{ab}$ 
by virtue of (3.6a) and (3.6b). It is also simple to check that 
$$
(M_{cd})_{ab\vpb} \ndg \gamma_{\ndg b \ndg \a \ndg \ad \vpb}
\ + \ (M_{cd})_{\a \b \vpb} \ndg \gamma_{a \ndg \b \ndg \ad \vpb}
\ + \ (M_{cd})_{\ad \bd } \ndg \gamma_{a \ndg \a \ndg \bd } \ = \ 0
\eqno(3.10) $$
so that the triality is invariant under $so(n)_\v$ provided that 
each space $\V$, $\S_\pm$ transforms appropriately.

Let us compare this with the construction of spin representations 
in general Euclidean dimension $n$. Spinors are commonly
described using complex vector spaces 
as opposed to the real spaces $\S_\pm$ we have been using,
but this is just a matter of notation: we can 
choose to write everything in real 
language if we wish, by regarding any complex vector space 
as a real space of twice the dimension. 
Once this has been done, the Dirac/Clifford construction 
for general $so(n)$ can always be expressed in terms of a map
$\gamma : \V {\times} \S_+ {\times} \S_- \rightarrow \R$
with $\V$ of dimension $n$, but with 
$\S_\pm$ now each having real dimension $N$, say.
The conditions (3.6a) and (3.6b) still apply and they ensure that 
the spin representations can be defined by (3.9), 
as before.

To translate this into more conventional language,
consider the maps 
$$ \r_a : \S_- \rightarrow \S_+ \ , \qquad 
\r_a{}^{\! \! \T} : \S_+ \rightarrow \S_- 
\eqno (3.11) $$ 
defined by (3.1b) for each basis vector in $\V$, 
as specified by the label $a$.
In matrix notation,
$$
(\r_a)_{\a \ad} \, = \, \gamma_{a \ndg \a \ndg \ad}
\eqno (3.12) $$
and the conditions (3.6a) and (3.6b) are equivalent to
$$\eqalignno{
\r_{a} \, \r_b{}^{\! \! \T} 
+ \ndg \, \r_{b} \, \r_a{}^{\! \! \T} \mdg & 
= \, 2 \ndg \delta_{ab} 
& (3.13{\rm a}) \cr
{\r_a}{}^{\! \! \T} \r_b
\, + \, {\r_b}{}^{\! \! \T} \r_a \, & 
= \, 2 \ndg \delta_{ab}
& (3.13{\rm b}) \cr
}
$$
(this also follows directly from (3.4)).
Standard Dirac matrices are then given by 
$$
\Gamma_a : \S \rightarrow \S \qquad {\rm with} \qquad 
\Gamma_a = \pmatrix{ 0 & \r_a \cr \r_a{}^{\! \! \T} & 0 \cr} 
\qquad 
{\rm on} 
\qquad 
\S = \S_+ \oplus \S_-
\eqno (3.14) $$
since (3.13) is clearly equivalent to 
$$
\{ \ndg \Gamma_a , \ndg \Gamma_b \ndg \} \, = \, 2 \ndg \delta_{ab} 
\eqno (3.15) $$
The representation of $so(n)_\v$ on $\S$ is reducible,
with generators $M_{cd} = \half \Gamma_{cd}$ 
having a block structure 
$$
\Gamma_{cd} \, \equiv \, 
\half ( \, \Gamma_c \Gamma_d - \Gamma_d \Gamma_c \, ) 
\, = \, 
\half \pmatrix{ 
\r_c \ndg \r_d{}^{\! \T} \mdg - \mdg \r_{d} \ndg \r_c{}^{\! \! \T} & 0 \cr
0 & \r_{c}{}^{\! \! \T} \r_d - \r_d{}^{\! \T} \r_{c} \cr }
\eqno(3.16) $$

From this more general perspective, we can derive a simple criterion
for the existence of a triality.
In discussing (3.7) above we noted that 
(3.6a) and (3.6b) are equivalent whenever
$\S_+$ and $\S_-$ have the same dimension, 
because $\r(u_\v)$ can then be expressed as a square matrix,
with a left inverse equal to its right inverse.
The same argument can be applied to 
$\r(u_\spm)$: these maps will correspond to square matrices,
and hence (3.6a) or (3.6b) will be equivalent to (3.6c), whenever 
$\S_\mp$ have the same dimension as $\V$. Such a matching of vector and
spinor dimensions, $N=n$, is therefore both necessary {\it and\/} 
sufficient for the existence of a triality since all
three conditions in (3.6) will be satisfied.\footnote{${}^{(1)}$}{
{\abs A similar argument was used in [5] in establishing a
correspondence with simple supersymmetric Yang-Mills.} }
We record this important conclusion in the 
following form.
\vskip 10pt

\noindent
{\bf Proposition:} If $\r_a$ with $a = 1, \ldots , n$ are
real $N{\times}N$ matrices 
obeying (3.13a) or (3.13b) (one implies the other) 
then (3.12) defines a triality $\gamma$ iff $N=n$.
\vskip 10pt

The possible values of $N$ which arise for each $n$ can  
be found by consulting one of the extensive treatments of
Clifford algebras and spin representations in the literature 
(e.g.~[6,9,10,17]) and these sources duly confirm that 
$N=n$ for precisely the four cases given in the Classification Theorem.
But it is also worth pointing out that only a rather small part of 
the whole edifice of Clifford technology is really 
required to reach this conclusion, and so it is just as well to state 
and prove what we need directly.
\vskip 10pt

\noindent
{\bf Lemma:} If $\Gamma_a$ with $a = 1, \ldots , 2 \ell$ are
$2N{\times}2N$ real symmetric matrices 
obeying (3.15) then $2^{\ell -1} \, | \, N$.
\vskip 5pt

\noindent
Proof: (3.15) implies that each matrix $\Gamma_{cd} = \half (\Gamma_c
\Gamma_d - \Gamma_d \Gamma_c )$
is real and antisymmetric with $(\Gamma_{cd})^2 = -1$. Its eigenvalues
are therefore $\pm i$, and a mutually commuting set such as 
$ \{  
\Gamma_{12} , \Gamma_{34} , \ldots , \Gamma_{2 \ell{-}1 \, 2 \ell} \} $
can be simultaneously diagonalised to produce a set of such
eigenvalues. 
But given any joint eigenvector, we can 
apply matrices $\Gamma_a$ to it and obtain new eigenvectors with 
any desired set of eigenvalues---applying $\Gamma_{1}$ or
$\Gamma_{2}$, for instance, gives a new eigenvector
with the eigenvalue of $\Gamma_{12}$ reversed but every other 
eigenvalue unchanged (using (3.15) again). 
All $2^\ell$ possible signs are therefore allowed 
for the eigenvalues, and since the eigenspaces have 
a common dimension (there are invertible maps between them)
$2N$ is a multiple of $2^{\ell}$. 
\vskip 10pt

\noindent
{\bf Corollary:} If $\gamma$ is a triality with $n > 1$, then 
$n$ is even and $2^{n / 2} \, |  \, 2 n$, implying $n = 2, 4$ or $8$.
\vskip 5pt

\noindent
Proof: Given a triality $\gamma$, there exist matrices $\r_a$ and 
$\Gamma_a$ with $N=n$, by the proposition.
If $n > 1$ there is at least one matrix 
$\Gamma_{ab}$ with a block form (3.16), and since  
$(\Gamma_{ab})^2 = -1$ the size $n$ of its blocks must be even.
The rest follows immediately from the lemma.
\vskip 10pt
 
Having shown that there are only four cases to consider,
existence can be established in a similarly 
elementary fashion. By appealing to the proposition once again,
it is sufficient to construct sets of matrices $\r_a$ 
obeying (3.13) with $N=n$, which can easily be done.
We will relegate further discussion to the appendix, however, 
which contains a fuller account of the relationship with 
the usual approach to Clifford algebra representations.

Treating $\V$ differently from $\S_\pm$ is helpful 
when comparing trialities with spinors 
in general dimensions, and we chose our notation with this in mind.
But there is nothing in the definition 
of a triality which distinguishes any one of 
the spaces $\V$, $\S_+$, $\S_-$ from the remaining two.
Moreover, equations such as (3.6)
are related to one another under permutations of these spaces,
and the arguments above involving $\r_a$, for instance, 
could equally well be given in terms of maps 
$\r_\a$ or $\r_{\ad}$ corresponding to
unit vectors in $\S_\pm$ instead of $\V$.
The idea that all three vector spaces appear in the
triality on exactly the same footing (our choice of notation
notwithstanding) is the key to everything which  
follows. We will investigate this more thoroughly 
in the next section by studying symmetries of trialities.
\vskip 50pt

\noindent
{\bf 4. Triality Symmetries}
\vskip 20pt

\noindent
Given any triality $\gamma$, there is a group
$O(n) \times O(n) \times O(n)$
which acts on $\V \times \S_+ \times \S_-$
and preserves the inner-product on each space. 
The {\bf triality group} is the subgroup of elements 
$(R_\v, R_\sp, R_\sm)$ under which $\gamma$ is
invariant\footnote{${}^{(2)}$}{{\abs 
This is the automorphism group, with isomorphism of trialities 
defined in (3.3); automorphism groups of isomorphic trialities are 
conjugate in $O(n) \! \times \! O(n) \! \times \! O(n)$.}
}
so that
$$ 
\gamma ( \, R_{\v} u_\v, \, R_\sp u_\sp, \, R_\sm u_\sm \, )
\ = \ \gamma (\, u_\v , \, u_\sp , \, u_\sm \, ) 
\eqno(4.1) $$
for all $u_\v$, $u_\sp$, $u_\sm$.
We will concentrate on 
the corresponding {\bf triality algebra}, denoted by
$$    
tri (n) \, \subset \, so(n) \oplus so(n) \oplus so(n)
\eqno(4.2) $$
and defined as the subalgebra of elements 
$$
T \ \ \leftrightarrow \ \ 
(\, M_\v , \, M_\sp, \, M_\sm ) \quad {\rm or} \quad
( \, M_{ab \vpb} , \, M_{\a \b \vpb} , \, M_{\ad \bd} \, ) 
\eqno(4.3{\rm a}) $$ 
which obey 
$$
\gamma ( \, M_\v u_\v , \, u_\sp, \, u_\sm )
\, + \,
\gamma ( \, u_\v , \, M_\sp u_\sp , \, u_\sm )
\, + \, 
\gamma ( \, u_\v , \, \ndg u_\sp, \, M_\sm u_\sm )
\, = \, 0
\eqno (4.3{\rm b}) $$
or, in components,
$$
M_{ab\vpb} \ndg \gamma_{\ndg b \ndg \a \ndg \ad \vpb}
\, + \, 
M_{\a \b \vpb} \ndg \gamma_{a \ndg \b \ndg \ad \vpb}
\, + \, 
M_{\ad \bd } \ndg \gamma_{a \ndg \a \ndg \bd } \, = \, 0
\eqno(4.3{\rm c}) $$
Note that when indices are written explicitly on any matrix $M$
they also serve to identify the space on which it acts, 
allowing us to omit the subscripts $\V$, $\pm$, just as we have chosen
to do when writing vectors in components.

The definition of $tri(n)$ as a subalgebra of 
$so(n) \oplus so(n) \oplus so(n)$ means that it inherits a 
positive-definite, invariant inner-product. It also 
means that the algebra comes equipped with 
representations on each of the spaces $\V$, $\S_+$, $\S_-$ 
given by the matrices $M$.
Let $\{ T_A \}$ be any orthonormal basis for $tri(n)$, with 
corresponding representation matrices 
$$
T_A \quad \leftrightarrow \quad 
( \, (M_A)_{ab}, \, (M_A)_{\a \b} , \, (M_A)_{\ad \bd} \, )
\eqno(4.4) $$
The overall scale of the inner-product can be fixed
in terms of these representations by imposing
$$
(T_A , T_B ) \ = \ k \, \Big \{ \, 
(M_A)_{ab \vpb}(M_B)_{ab \vpb}
\, + \,
(M_A)_{\a \b \vpb} (M_B)_{\a \b \vpb} 
\, + \, 
(M_A)_{\ad \bd} (M_B)_{\ad \bd}
\, \Big \} 
\ = \ \delta_{AB}
\eqno (4.5) $$ 
The value of the coefficient $k$ is a detail
to which we will return below.

Our earlier result (3.10) can now be expressed by saying that  
$$so(n)_\v = \wedgesq  \V \quad {\rm with~basis} \quad  
\{ \ndg T_{cd} \ndg \}
\eqno(4.6{\rm a}) $$
and representations (3.8) and (3.9) is a subalgebra of $tri(n)$.
But from the permutation symmetry of (3.6) it is clear that we 
can equally well construct other subalgebras 
$$so(n)_\spm = \wedgesq \S_\pm
\quad {\rm with~ bases} \quad  
\{ \ndg T_{\c \d \vpb} \ndg \} \ \ {\rm or} \ \  \{ \ndg T_{\cd \dd} \ndg \}
\eqno(4.6{\rm b}) $$ 
simply by interchanging the roles of the three spaces. 
To carry this out explicitly we introduce a little more notation,
defining 
$$\eqalignno{
(\gamma_{ab})_{\a \b \vpb} \, \, \equiv \, \ndg \ndg 
(\gamma_{\a \b})_{a b \vpb} \, \, & \equiv \,  
\half (\, 
\gamma_{ a \ndg \a \ndg \ad \vpb} \gamma_{ \ndg b \ndg \b \ndg \ad \vpb}
- 
\gamma_{ \ndg b \ndg \a \ndg \ad \vpb}  \gamma_{ a \ndg \b \ndg \ad
\vpb} 
\, ) 
\qquad & (4.7{\rm a}) \cr
(\gamma_{\ad \bd})_{a b \vpb} 
\, \, \equiv \, \ndg \ndg 
(\gamma_{ab})_{\ad \bd} 
\, \, & \equiv \,  
\half (\, 
\gamma_{ a \ndg \a \ndg \ad \vpb} \gamma_{ \ndg b \ndg \a \ndg \bd \vpb}
- \gamma_{ \ndg b \ndg \a \ndg \ad \vpb}  \gamma_{ a \ndg \a \ndg \bd
  \vpb} 
\, ) 
\qquad & (4.7{\rm b}) \cr
(\gamma_{\a \b})_{\ad \bd} \, \equiv \, 
(\gamma_{\ad \bd})_{\a \b \vpb} \, & \equiv \,  
\half (\, \gamma_{ a \ndg \a \ndg \ad \vpb} \gamma_{ a \ndg \b \ndg \bd \vpb}
- \gamma_{ a \ndg \b \ndg \ad \vpb}  \gamma_{ a \ndg \a \ndg \bd \vpb} \, ) 
\qquad & (4.7{\rm c}) \cr
}$$
These antisymmetrised combinations of $\gamma$ components generalise 
those in (3.9) which were used to introduce 
spin representations of (4.6a); the results can be summarised 
$$
so(n)_\v \ : 
\quad 
(M_{cd})_{ab}  = 
\delta_{ac} \ndg \delta_{bd} - \delta_{ad} \ndg \delta_{bc} 
\ , \ \ \qquad
(M_{cd})_{\a \b} = \half (\gamma_{cd})_{\a \b \vpb} 
\ , \qquad
(M_{cd})_{\ad \bd} = \half (\gamma_{cd})_{\ad \bd} 
\qquad \eqno(4.8{\rm a}) $$
acting on $\V$, $\S_+$, $\S_-$.
Now we define, similarly, the algebras (4.6b) and their
representations 
$$
so(n)_\sp \ : 
\quad 
(M_{\c\d})_{\a\b} = 
\delta_{\a\c} \ndg \delta_{\b\d} - \delta_{\a\d} \ndg \delta_{\b \c} 
\ , \qquad
(M_{\c\d})_{\ad \bd} = \half (\gamma_{\c\d})_{\ad \bd} 
\ , \qquad
(M_{\c\d})_{a b \vpb} = \half (\gamma_{\c\d})_{a b \vpb} 
\qquad \eqno(4.8{\rm b}) $$
acting on $\S_+$, $\S_-$, $\V$ respectively; and  
$$
so(n)_\sm \ : 
\quad 
(M_{\cd\dd})_{\ad\bd} = 
\delta_{\ad\cd \vpb} \ndg \delta_{\bd\dd} - \delta_{\ad\dd} \ndg 
\delta_{\bd\cd} 
\ , \qquad
(M_{\cd\dd})_{a b \vpb} = \half (\gamma_{\cd\dd})_{a b \vpb} 
\ , \qquad
(M_{\cd\dd})_{\a \b \vpb} = \half (\gamma_{\cd\dd})_{\a \b \vpb} 
\qquad \eqno(4.8{\rm c})$$
acting on $\S_-$, $\V$, $\S_+$ respectively. 
The invariance of the triality under any one of 
these $so(n)$ algebras is easy to verify; on the other 
hand, the permutation symmetry of (3.6) 
implies that $\gamma$ must be invariant under $so(n)_\spm$
once we know it is invariant under $so(n)_\v$.

The algebras and representations in (4.8) can also be described in a
basis-independent way, of course. Given $u$ and $w$ belonging to 
one of the spaces $\V$, $\S_+$, $\S_-$, we can regard $u \wedge w$ 
as acting in the fundamental representation of the appropriate 
$so(n)$ algebra, as discussed in section 2; the relations (3.4) 
then ensure that we also have corresponding actions on the other 
two spaces given by maps 
$$\eqalign{ 
& \quarter( \ \r (u ) \, \r ( w )^{\mdg \T} - \, \r (w ) \, \r ( u
)^{\mdg \T} \, ) 
\cr
& \quarter( \ \r (u )^{\mdg \T} \r ( w ) \, - \, \r (w )^{\mdg \T} 
\r ( u ) \ ) 
\cr
}
\eqno (4.9) $$
The component versions (4.8), with the accompanying notation introduced 
in (4.7), prove to be extremely useful and versatile, however, 
so we shall continue to deal mainly with these forms.
To help keep the meanings clear despite the 
proliferation of indices attached to each $M$ or $\gamma$,
observe that in (4.8) pairs of indices inside brackets label
Lie algebra generators, while those outside refer
to the vector spaces on which the generators act. 
We also note for future reference that the 
defining relations for a triality in (3.6) can now
be re-expressed 
$$\eqalignno{
\gamma_{ a \ndg \a \ndg \ad \vpb} \ndg \gamma_{ \ndg b \ndg \b \ndg \ad \vpb}
\ & = \ 
\delta_{ab \vpb} \ndg \delta_{\a \b \vpb} \, \, + \, \ndg 
(\gamma_{ab})_{\a \b \vpb} 
& (4.10{\rm a}) \cr
\gamma_{ a \ndg \a \ndg \ad \vpb} \ndg \gamma_{ \ndg b \ndg \a \ndg \bd \vpb}
\ & = \ 
\delta_{ab \vpb} \ndg \delta_{\ad \bd \vpb} \, \, + \, \ndg
(\gamma_{\ad \bd})_{a b} 
& (4.10{\rm b}) \cr
\gamma_{ a \ndg \a \ndg \ad \vpb} \ndg \gamma_{ a \ndg \b \ndg \bd \vpb}
\ & = \ 
\delta_{\a \b \vpb} \ndg \delta_{\ad \bd \vpb} \, + \, 
(\gamma_{\a \b})_{\ad \bd} 
& (4.10{\rm c}) \cr
}$$

There is still a technical detail to clear up before we move 
on. In section 2 we fixed normalisation conventions for 
$so(n)$ and its inner-product, expressing these in terms of 
the fundamental representation by equations 
(2.17) through to (2.21). We will insist on exactly the same 
conventions for $so(n)_\v$ and $so(n)_\spm$, and 
by comparing their fundamental representations on $\V$ and $\S_\pm$ 
in (4.8) with (2.21), we see that this corresponds to the 
bases in (4.6) being orthonormal. But we also want these
inner-products to arise as the restrictions of the 
inner-product on $tri(n)$ given in (4.5). 
The normalisations work out correctly provided we choose 
$$
k = (\half n + 2)^{-1} 
\eqno (4.11)$$
as can be checked from (4.8).

It remains to actually determine $tri(n)$ and to understand 
how its three $so(n)$ subalgebras interlock. 
Since there are only four cases to consider, 
it is tempting just to analyse each of them in turn,
but there are benefits to delaying such a case-by-case 
approach for as long as possible.
We will now derive some simple results 
on the structure of $tri(n)$, for any $n$, 
which will eventually be crucial for our uniform 
construction of the algebras in the magic square.
We will also see, in the next section, how these  
results can be applied to reveal the specific 
properties of each triality rather easily.

A natural way to investigate $tri(n)$ is to 
take each $so(n)$ subalgebra 
and ask what is left over.
Considering $so(n)_\v$, for instance, we can plainly express 
{\it any} antisymmetric matrix acting on $\V$ as a linear combination of the 
generators in (4.8a). So, by linearity, it suffices to 
consider the subalgebra of $tri(n)$ which fixes all elements in $\V$,
consisting of what might reasonably be called {\it internal}
transformations (as far as $\V$ is concerned). 
We will therefore define $int(n)_\v$ to be the subalgebra of 
$tri(n)$ with $M_\v = 0$; or 
$$
T_A \ \ \ {\rm is~a~generator~of} \ \ \ int(n)_\v 
\quad \iff \quad (M_A)_{ab \vpb} = 0
\eqno(4.12{\rm a}) $$
Similarly, we define $int(n)_\spm$ as the subalgebras 
of $tri(n)$ which fix all elements in $\S_\pm$, so that 
$M_\pm = 0$; or  
$$
T_A \quad {\rm is~a~generator~of} \quad int(n)_\spm 
\quad \iff \quad 
(M_A)_{\a\b \vpb} = 0 ~~{\rm or}~~(M_A)_{\ad\bd} = 0 
\eqno(4.12{\rm b}) $$
From these definitions we can deduce the following.
\vskip 10pt

\noindent
{\bf Proposition}: 
As orthogonal direct sums of Lie algebras
$$\eqalign{
tri (n) \ & = \ so(n)_\v \oplus int(n)_\v \cr
\ & = \ so(n)_\sp \oplus int(n)_\sp \cr
\ &= \ so(n)_\sm \oplus int(n)_\sm \cr
}\eqno(4.13)$$
In addition $int(n)_\v$, $int(n)_\sp$, $int(n)_\sm$ 
are commuting subalgebras with trivial pairwise intersections 
$$
int(n)_\v \, \cap \, int(n)_\spm \ = \ int(n)_\sp \, \cap \, int (n)_\sm \ =
\ \{ 0 \}
\eqno(4.14) $$
\vskip 5pt

\noindent
Proof: We can fix attention on $\V$, with the understanding that 
similar arguments will also apply
to $\S_\pm$. We have already noted that 
any element of $tri(n)$ is a sum of elements in $so(n)_\v$ 
and $int(n)_\v$ and their intersection is clearly trivial. 
Now if $( 0, \lambda_{\a \b \vpb} , \lambda_{\ad \bd})$ belongs 
to $int(n)_\v$ then (4.3) states that
$$
\lambda_{\a \ndg \b \ndg \vpb} \, \gamma_{a \ndg \b \ndg \ad \vpb}
\, + \, \lambda_{\ad \bd } \, \gamma_{a \a \bd } \, = \, 0
\eqno (4.15) $$
and it is easy to deduce
(by contracting with $\gamma_{b \c \ad}$ or 
$\gamma_{b \a \cd}$)  
that $\lambda_{\a \b \vpb}$ and $(\gamma_{ab})_{\a \b \vpb}$ commute,
as do $\lambda_{\ad \bd}$ and $(\gamma_{ab})_{\ad \bd}$.
Referring to (4.8a), this means that $so(n)_\v$ and $int(n)_\v$ 
are commuting subalgebras, and so $tri(n)$ is indeed their direct sum. 
A further consequence of (4.15), which can be obtained in a similar
fashion, is 
$$
\lambda_{\a \b \vpb} \ndg (\gamma_{ab})_{\a \b \vpb}  \, + \, 
\lambda_{\ad \bd} \ndg (\gamma_{ab})_{\ad \bd \vpb} \, = \, 0
\eqno(4.16) $$
and so the subalgebras are orthogonal with respect to the 
inner-product in (4.5).
The remaining assertions in the proposition are 
also simple consequences of the definitions; for example,  
the intersection of any pair of subalgebras is trivial since 
if two of the matrices $M$ vanish in (4.3) then 
so must the third.
\vskip 10pt

\noindent
{\bf Corollary:} The subalgebras 
$so(n)_\v$, $so(n)_\sp$, $so(n)_\sm$ arise  
as projections of $tri(n)$, with their generators (4.6) 
given in terms of the corresponding representation matrices by
$$
T_{ab \vpb} = (M_A)_{ab \vpb} \ndg T_A \ , \qquad 
T_{\a \b \vpb} = (M_A)_{\a \b \vpb} \ndg T_A \ , \qquad 
T_{\ad \bd} = (M_A)_{\ad \bd} \ndg T_A
\eqno (4.17) $$
\vskip 5pt

\noindent
Proof: Compare with the treatment of $so(n)$ in section 2, 
leading to (2.19).
Since we have been careful to adopt the same normalisation 
conventions for all copies of $so(n)$, the only difference 
lies in the nature of the sum over $A$. 
In (4.17), as throughout this section, $A$ labels 
a basis for $tri(n)$, and this is strictly larger than each $so(n)$ subalgebra 
if $int(n)$ is non-trivial. But the orthogonal decompositions in 
the proposition above
imply that each sum in (4.17) can be split into a contribution from 
the appropriate $so(n)$ subalgebra and another from the companion 
$int(n)$ factor, and the definitions (4.12) imply that the 
latter contributions vanish. 
The results therefore follow, just as in section 2.
\vskip 10pt

It is instructive to 
check this by showing it is consistent with the results in section 2 
in a slightly different way.
Taking the appropriate representations of each generator $T$ in 
(4.17) and then using (4.8) gives
$$\eqalign{
(M_A)_{ab \vpb}(M_A)_{cd \vpb} 
\ = \ \, 
(M_{ab})_{cd} 
& \ = \ \delta_{a c} \delta_{bd} - \delta_{ad} \delta_{bc} 
\cr
(M_A)_{\a\b \vpb}(M_A)_{\c \d \vpb} 
\ = \,
(M_{\a\b})_{\c\d} 
& \ = \ \delta_{\a \c} \delta_{\b \d} - \delta_{\a\d} \delta_{\b\c} 
\cr
(M_A)_{\ad \bd \vpb}(M_A)_{\cd \dd \vpb} 
\ = \, 
(M_{\ad\bd})_{\cd\dd} 
& \ = \ \delta_{\ad \cd \vpb} \delta_{\bd\dd} - \delta_{\ad\dd} \delta_{\bd\cd} 
\cr
}
\eqno (4.18) $$
But we can also arrive at the end results directly, by comparison 
with (2.21). The sums here have a larger range, in general, 
since $A$ runs over a basis for $tri(n)$, but once again the sum can be
split into a contribution from an $so(n)$ subalgebra 
and another from its orthogonal complement $int(n)$, with the latter 
making no contribution. Other, similar relations can also be deduced from 
(4.17) by taking representations of $T$ other than the fundamental, 
and we will find this idea useful later.

Returning to the proposition itself, there is also another 
way to understand the direct sums in (4.13), based on the 
following general observation.
If a Lie algebra with an invariant positive-definite 
inner-product has a unitary representation $\rho$, then the 
algebra is isomorphic to ${\rm Ker} \rho \oplus {\rm Im} \rho$.
We can apply this to $tri(n)$ and each
of its representations on $\V$, $\S_\pm$: it is easy to see that 
the image of each representation is isomorphic to the corresponding 
copy of $so(n)$, while the
kernel is the relevant $int(n)$ algebra. Hence, we obtain the three 
orthogonal decompositions in (4.13).

The dominant theme in our discussion of triality symmetries has been
the permutation symmetry which is manifest in all key relations, from 
the definitions (3.1) or (3.6), through the construction of the 
algebras (4.6) and (4.8), to the proposition above and the 
decompositions in (4.13).
This symmetry entails nothing more than 
the ability to interchange, wholesale, the roles of the 
underlying spaces $\V$, $\S_+$, $\S_-$ 
in everything we do, whether in constructions or proofs.
There is a much more sophisticated and elegant version of this idea,
which the concept of a triality was specifically designed to capture, 
and which we will present in the form of a theorem.
This will be the logical culmination of our discussion of symmetries, 
and it provides a new perspective on the results we have obtained so
far. Nevertheless, these earlier conclusions are quite sufficient,
in themselves, to allow us to construct the magic square, 
without appealing directly to the more sophisticated result 
which now follows.

To begin with, we must prepare the ground.
\vskip 5pt

\noindent
{\bf Lemma:} Given a triality $\gamma$, let $e_\v$ be any unit vector in 
$\V$ and consider the associated orthogonal maps 
$$
\tau (\e) : \V \rightarrow \V \ , \qquad 
\r (\e) : \S_- \rightarrow \S_+ \ , \qquad 
\r (\e)^\T : \S_+ \rightarrow \S_-
\eqno(4.19) $$
with $\r(e_\v)$ defined in (3.1) and $\tau (\e)$ a standard
reflection
$$
\tau (e_\v ) u_\v = u_\v - 2 (u_\v \cdot e_\v ) e_\v \qquad 
{\rm or} \qquad 
\tau (e_\v)_{ab} \, = \, \delta_{ab} \, - \, 2 e_a e_b 
\eqno(4.20) $$
the latter equation giving its matrix form. Then
$$
\gamma ( \, \tau(\e)u_\v \, , \, \r (\e) u_\sm , \, \r (\e)^\T u_\sp \, )
\ = \
- \gamma ( \, u_\v  , \, u_\sp , \, u_\sm \, )
\eqno(4.21) $$ 
for all $u_\v$, $u_\sp$, $u_\sm$.
\vskip 5pt

\noindent
Proof: This is easily confirmed by a short calculation to unwind the
definitions, either in components or using 
a basis-independent approach.  
\vskip 5pt

\noindent
Since composing reflections gives rotations, the lemma provides 
an alternative way to investigate the invariance of $\gamma$, and one 
which allows the study the triality {\it group\/} rather than just 
its algebra. From the result as stated, we can deduce invariance under 
a subgroup corresponding to $so(n)_\v$, but by considering 
reflections on $\S_\pm$ and associated maps between the remaining 
pairs of spaces we can equally well construct the groups
corresponding to $so(n)_\spm$. Our concern here, however, is to put
the lemma to a different use.
\vskip 10pt

\noindent
{\bf Permutation Theorem}: For $n > 1$, 
$tri(n)$ has a permutation group $S_3$ of outer automorphisms 
which acts on its 
representations $\V$, $\S_+$, $\S_-$, or equivalently permutes its 
subalgebras $so(n)_\v$, $so(n)_\sp$, $so(n)_\sm$ and also 
$int(n)_\v$, $int(n)_\sp$, $int(n)_\sm$.
\vskip 5pt

\noindent
Proof: Outer automorphisms are equivalence classes of
automorphisms (see e.g.~[1,2]) and we can define 
representatives in each class. It suffices to 
find an automorphism which exchanges any pair of spaces,
and so we consider $\S_+$ and $\S_-$.
Let $\tau$ and $\r$ be the maps defined by some unit vector $e_\v$
(the dependence on which we now choose to suppress, to simplify 
notation) as in the lemma above. 
It is straightforward to check that (4.21) implies 
that
$$
(\, M_\v , \, M_\sp , \, M_\sm \,  ) 
\  \mapsto  \ 
( \, \tau M_\v \tau \, , 
\, \r M_\sm \r^\T \, ,  
\, \r^\T \! M_\sp \r \, ) 
\eqno(4.22)$$
is an automorphism of $tri(n)$. It certainly exchanges the
representations on $\S_\pm$ up to similarity, and 
different choices of the unit vector $e_\v$ are related by rotations 
in $\V$ and correspond to the same equivalence class of automorphisms. 
\vskip 10pt

In looking back over our general treatment of trialities, we can 
underscore the close links with ideas in 
theoretical physics. Although we are dealing here with Euclidean
signature, the construction of spin representations of $so(n)$ follows 
the same route as any elementary introduction 
to the Dirac equation. The invariance of the triality (3.10) is 
also obtained in just the same way that one checks Lorentz-invariance
of the Dirac lagrangian, and the lemma leading to the Permutation
Theorem can be viewed as a discussion of parity for spinors. 
At a slightly less elementary level, the $int(n)$ algebras arise 
in the same fashion as $R$-symmetries in supersymmetric
theories, and our proof 
of the Classification Theorem has 
strong links with the treatment of 
supersymmetric Yang-Mills (or classical superstrings) given in [5].
\vskip 50pt

\noindent
{\bf 5. Trialities Case by Case}
\vskip 20pt

\noindent
We will now use the general results of sections 3 and 4
to derive the distinctive properties of each 
of the four trialities. Our strategy will be to 
consider 
$$
tri(n) \, \subset \, so(n) \oplus so(n) \oplus so(n)
\eqno (5.1) $$
in conjunction with the results (4.13) and (4.14) of the proposition
in the last section.
Together these impose strong constraints, but with 
different consequences for each value of $n$. 
As far as our principal aim of constructing the magic square is concerned,
none of the results in this section are needed to understand how or why 
this works; they are relevant to 
identifying the algebras which emerge from the construction, however.

We begin with the case $n=8$. Since $so(8)$
is a simple Lie algebra, we can deduce immediately from (4.13) and
(4.14) that 
$$
tri(8) = so(8)_\v = so(8)_\sp = so(8)_\sm \  
\qquad {\rm with } \qquad 
\ \ \ int(8) \ \ \ {\rm trivial} 
\eqno (5.2) $$
This is a striking result, sometimes referred to as 
{\it Principle of Triality} (see [17] for references to the 
earlier literature).
It states that if the action of $tri(8)$
is given on any one of $\V$, $\S_+$ or $\S_-$, then its action on 
the other spaces is determined by (4.3).
(We are always working locally, at the level of algebras; the
statement for the corresponding groups must take account of finite, discrete
ambiguities associated with their global structure.) 

Since {\it any} two copies of 
$so(8)$ are isomorphic, it is important to realise that
the equalities above tell us far more:
each of these $so(8)$ algebras consists of precisely the same 
set of transformations, despite their differing
actions on $\V$, $\S_+$, $\S_-$. 
It must be possible, therefore, to relate the orthonormal bases 
defined in (4.6) and (4.8), and this has effectively been achieved 
already by the corollary of the last section. To be more specific, 
if we replace the index $A$ in
(4.17) by various kinds of antisymmetric pairs, we can obtain,
for instance,
$$
T_{\a \b \vpb} = \half (M_{ab \vpb})_{\a \b \vpb} \ndg T_{ab \vpb} \ , \qquad 
T_{\ad \bd} = \half (M_{\a \b \vpb})_{\ad \bd} \ndg T_{\a \b \vpb} \ , \qquad 
T_{ab} = \half (M_{\ad \bd})_{ab \vpb} \ndg T_{\ad \bd} 
\eqno (5.3) $$
(recall the convention explained in section
2, according to which a factor $\half$ naturally accompanies 
such sums over antisymmetric pairs).
The representation matrices $M$ have now taken on a new interpretation
(via implicit dualisations, yet again)---they provide the changes of
base which allow us to identify 
$$
\wedgesq \V = \wedgesq \S_+ = \wedgesq \S_- \qquad {\rm for} \ \ \ \ n = 8
\eqno (5.4) $$

The Permutation Theorem is manifest in the symmetry of the 
Dynkin diagram of $so(8)$, with $S_3$ acting 
on the outer nodes corresponding to the representations $\V$, $\S_+$,
$\S_-$.
\vskip 20pt

\hbox{ \kern 95pt \kern 100pt 
\lower 16pt \hbox{\epsfxsize=60pt \epsfbox{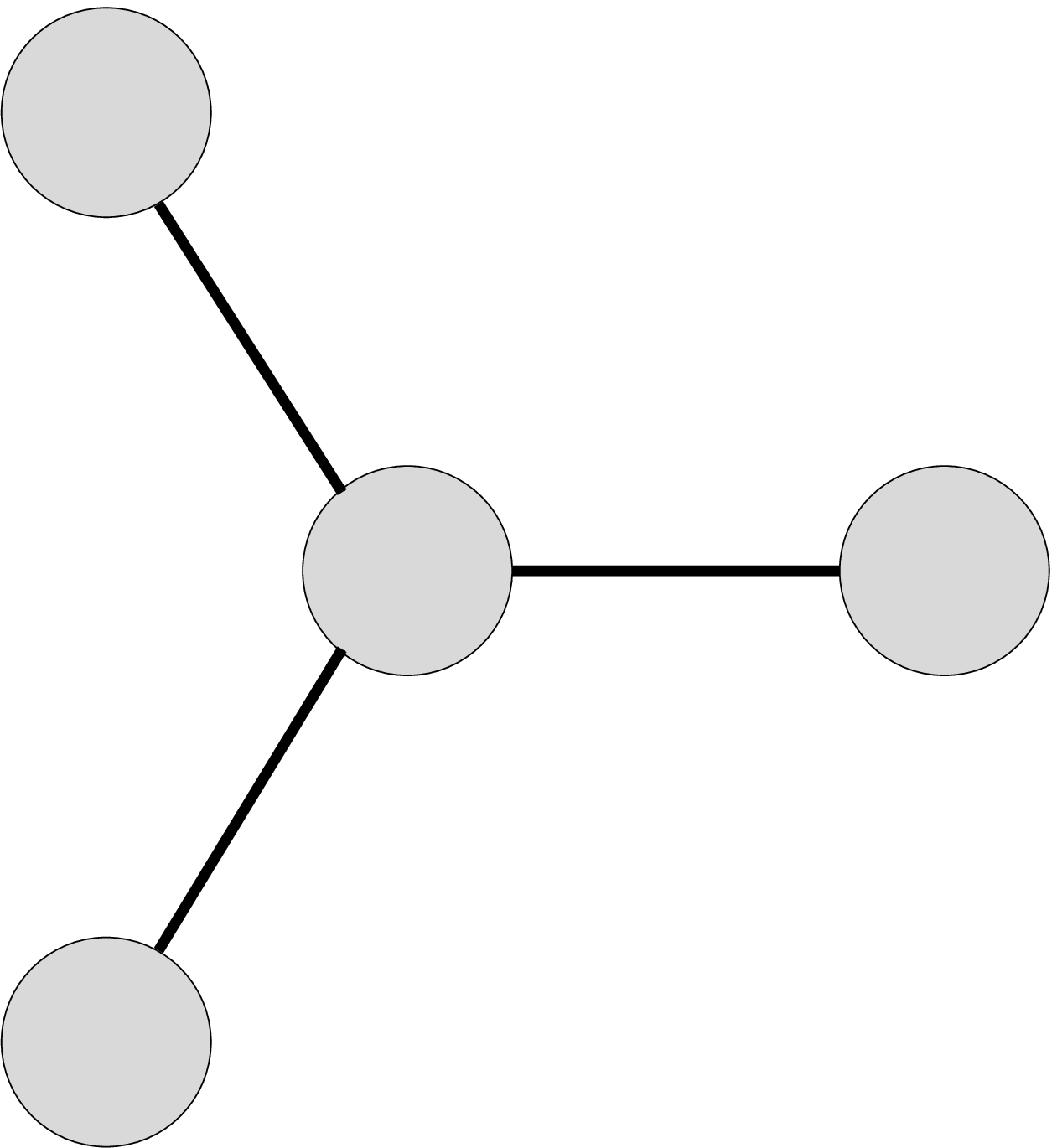}} } \vskip 15pt

\centerline{Figure 1: The Dynkin diagram of $tri(8) = so(8)$}
\vskip 15pt

\noindent
The central node corresponds to the adjoint
representation, which is exactly what appears in (5.4). 
As the vector and spinor representation are 
permuted, the adjoint representation is inert, but only up to the 
changes of base, as in (5.3).

Of the remaining trialities,
the case with $n=1$ is rather degenerate. With $\V$ and $\S_\pm$
one-dimensional, we can identify each of them
with $\R$ and take the triality map to be 
$$
\gamma (x, y , z) \ = \ xyz
\eqno (5.5) $$ 
without loss of generality.
The Lie algebras $tri(n)$, $so(n)$ and $int(n)$ are all   
trivial when $n=1$, of course, and so this case does not 
even qualify for inclusion in the Permutation
Theorem (although this could be adapted to apply to the 
discrete triality {\it group} instead [1]).
The cases $n=2$ and $n=4$ are more interesting. 
They each have $tri(n)$ strictly larger than $so(n)$, and the presence of 
non-trivial $int(n)$ factors is essential in understanding 
both the existence of the triality and its 
$S_3$ permutation symmetry. We will consider each case in turn.

The major difference in moving from $n=8$ to $n=4$ is that we 
now have a semi-simple algebra
$$
so(4) = su(2) \oplus su(2) \qquad {\rm with} \qquad int (4) = su(2)
\eqno (5.6) $$
This is the only possibility compatible with (5.1), (4.13) and (4.14);
more precisely, we can write 
$$
tri(4) \ = \ su(2)_\sp \oplus su(2)_\sm \oplus su(2)_\v 
\eqno (5.7) $$
where the factors on the right are defined so that  
$$
\eqalign{
so(4)_\v & = \ su(2)_\sp \oplus su(2)_\sm \ , \qquad int(4)_\v = su(2)_\v \cr
so(4)_\spm & = \ su(2)_\smp \oplus su(2)_\v \ , 
\qquad int(4)_\spm = su(2)_\spm \cr
\cr
} \eqno (5.8) $$
As representations of $tri(4)$ we have 
$$\eqalign{
\V \ & = \, ({\bf 2}_\sp , {\bf 2} _\sm , {\bf 1}_\v) 
\cr
\S_+ \! & = \, ({\bf 1}_\sp , {\bf 2}_\sm , {\bf 2}_\v) 
\cr
\S_- \! & = \, ({\bf 2}_\sp , {\bf 1}_\sm , {\bf 2}_\v) 
\cr
} \eqno (5.9) $$
where ${\bf 2}$ and ${\bf 1}$ denote the fundamental and trivial
representations of each $su(2)$, as usual. 
The Dynkin diagram of $tri(4)$ is disconnected, 
but still exhibits the $S_3$ symmetry required 
by the Permutation Theorem. 
\vskip 20pt

\hbox{ \kern 100pt \kern 100pt 
\lower 16pt \hbox{\epsfxsize=50pt \epsfbox{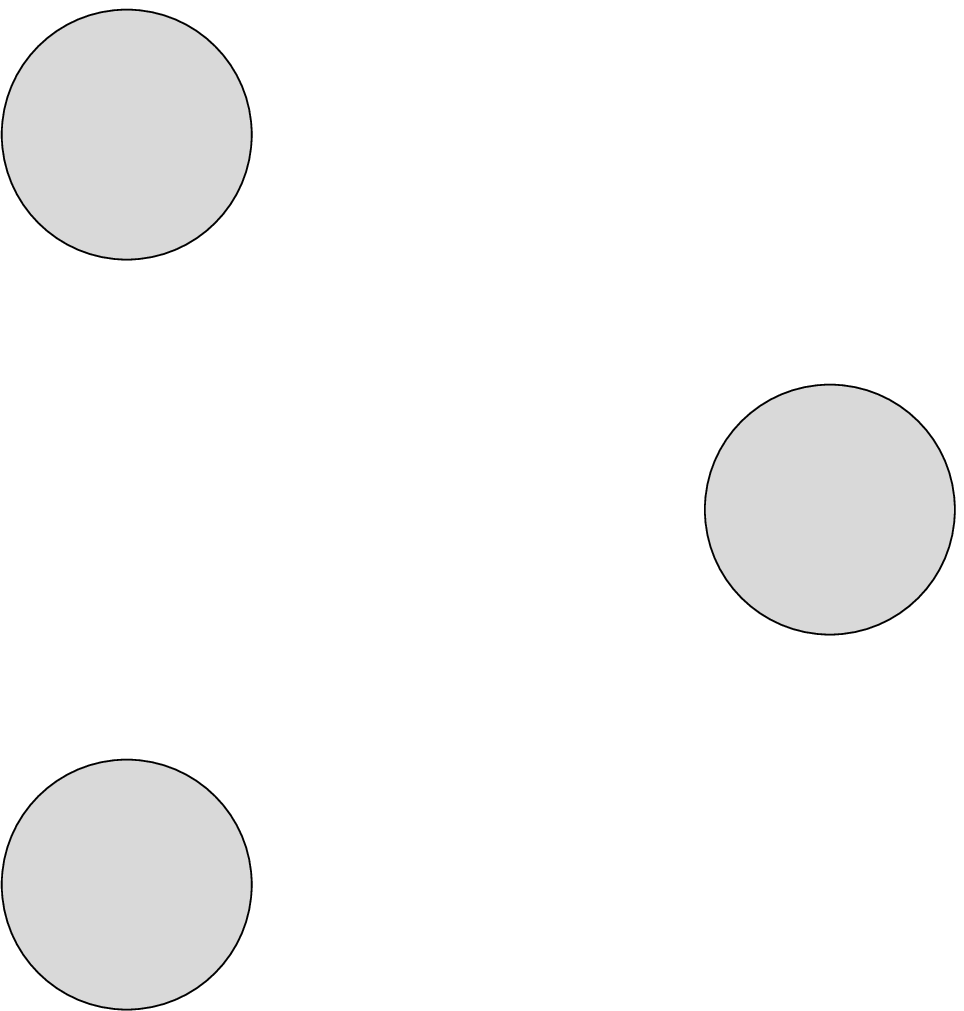}} } \vskip 15pt

\centerline{Figure 2: The Dynkin diagram of $tri(4) = su(2) \oplus
su(2) \oplus su(2)$}
\vskip 15pt

\noindent
Its nodes correspond to the representations ${\bf 2}_\v$ and 
${\bf 2}_\spm$, so pairs of nodes correspond to $\V$ and $\S_\pm$.

To elaborate on the various statements above, let us first recall
that $so(4) = su(2) \oplus su(2)$ can be understood as 
a decomposition of the space of antisymmetric $4{\times}4$ 
matrices into self-dual and anti-self-dual (or $\pm$-self-dual) subspaces. 
Consider, for definiteness, a
matrix $M_{ab}$ acting on $\V$, and define this to 
be $\pm$-self-dual iff 
$$
M_{ab} = \pm \half \eps_{abcd} M_{cd} 
\eqno (5.10)
$$
where $\eps_{abcd}$ is the usual alternating symbol in four
dimensions. Any antisymmetric matrix can be written as a sum of its 
self-dual and anti-self-dual parts, 
and it is easy to check that the $\pm$-self-dual 
subspaces are each closed under commutation and mutually commuting.
This can also be expressed 
$$
so(4)_\v \ = \ \wedgesq \V \ = \ 
\wedge^2_+ \V \oplus \wedge^2_- \V
\eqno (5.11) $$
corresponding to the generators in the vector representation 
being decomposed 
$$
(M_{cd})_{ab} \ = \ 
\half \left \{ \delta_{ac} \delta_{bd} - \delta_{ad} \delta_{bc} 
+ \half \eps_{abcd} \right \}
\ + \ 
\half \left \{ \delta_{ac} \delta_{bd} - \delta_{ad} \delta_{bc} 
-\half \eps_{abcd} \right \}
\eqno (5.12) $$
Each of the quantities in braces above are 
$\pm$-self-dual in {\it both} sets of indices $ab$ and $cd$.
The other subalgebras can obviously be decomposed in a similar
fashion
$$
so(4)_\spm = \wedgesq \S_\pm = 
\wedge^2_\sp \S_\pm \oplus \wedge^2_\sm \S_\pm
\eqno (5.13) $$

Now consider the implications for the representations of 
$so(4)_\v$ on $\S_\pm$ (and, similarly, with spaces permuted). 
A useful elementary fact is that any 
antisymmetric matrix whose square is a non-zero multiple of the
identity must be either self-dual or anti-self-dual.
The symbols such as 
$$
(\gamma_{ab \vpb})_{\a \b \vpb}
\  \qquad 
(\gamma_{\a \b \vpb})_{\ad \bd}
\  \qquad 
(\gamma_{\ad \bd})_{a b \vpb}
\eqno (5.14) $$
can be regarded as matrices in {\it either\/} pair of indices,
and (3.6) implies that, as such, they always square to $-1$ 
(with the other pair of indices fixed throughout). 
They must, therefore, be either self-dual or anti-self-dual 
in {\it every\/} pair of indices. Which possibility occurs in
to some extent a matter of convention, because the conditions 
for $\pm$-self-duality are reversed under a change in
orientation (so under any change of base with 
negative determinant).
But we do not have complete freedom of choice, 
since the algebra (3.6) also implies that 
if $(\gamma_{ab})_{\a \b \vpb}$ is self-dual in $\a \b$, then 
$(\gamma_{ab})_{\ad \bd}$ is anti-self-dual in $\ad \bd$, and vice
versa. This means that in the decomposition (5.11) the 
subalgebras $\wedge^2_\pm \V$ each act non-trivially on just one of 
the spaces $\S_\pm$, which is of course a familiar feature of the spin
representations for $n=4$.
   
To fix the freedom consistently while 
preserving as much symmetry as possible, we will adopt
the convention that the $\gamma$ symbols as written in (5.14) are 
anti-self-dual on each pair of indices inside the brackets 
and self-dual on each pair of indices outside. This can 
always be achieved by re-ordering the bases in $\V$ and $\S_\pm$, 
if necessary, and once done it allows us to identify the 
$\pm$-self-dual parts of different copies of 
$so(4)$ within $tri(4)$, according to
$$
T_{\a \b \vpb} = \quarter (\gamma_{ab \vpb})_{\a \b \vpb} T_{ab \vpb} 
\ , \qquad 
T_{\ad \bd} = \quarter (\gamma_{\a \b \vpb})_{\ad \bd} T_{\a \b \vpb} 
\ , \qquad 
T_{ab \vpb} = \quarter (\gamma_{\ad \bd})_{ab \vpb} T_{\ad \bd} 
\eqno (5.15) $$
These relations follow from (4.17), just like (5.3) for $so(8)$,
but we have written things in terms of $\gamma$ symbols to emphasise the 
consequences of our choice of convention; each relation is invertible,
given the $\pm$-duality properties of the index pairs.
Taking these identifications into account, we can now
define $su(2)$ algebras:
$$\eqalign{
su(2)_\v & \equiv \wedge_\sm^2 \S_+ = \wedge_\sp^2 \S_- 
\cr
su(2)_\sp & \equiv \, \wedge_\sp^2 \V \, = \, \wedge_\sm^2 \S_- 
\cr
su(2)_\sm & \equiv \, \wedge_\sm^2 \V \, = \, \wedge_\sp^2 \S_+ 
\cr
}
\eqno (5.16) $$
and then (5.11) and (5.13) produce (5.7) and (5.8).

The last triality to consider has $n=2$. This is rather elementary
since all the algebras concerned are abelian and we 
can identify 
$$ so(2) = int(2) = u(1) \qquad {\rm and} \qquad 
tri(2) = u(1) \oplus u(1) 
\eqno (5.17) $$
It is easiest to describe everything 
embedded in $so(2) \oplus so(2) \oplus so(2) = 
u(1) \oplus u(1) \oplus u(1)$ acting on 
$\V \times \S_+ \times \S_-$, 
so each generator can be specified 
by a triple of real numbers or weights.
The subalgebra whose weights add to zero is exactly $tri(2)$,
and the various subalgebras it contains correspond to 
weights as follows
$$\eqalign{
so(2)_\v \sim (2, -1, -1) \ , & \qquad int(2)_\v \sim (0, 1, -1) \cr 
so(2)_\sp \sim (-1, 2, -1) \ , & \qquad int(2)_\sp \sim (-1, 0, 1) \cr 
so(2)_\sm \sim (-1, -1, 2) \ , & \qquad int(2)_\sm \sim (1, -1, 0) \cr 
}
\eqno (5.18) $$

Finally, note that the appearance of non-trivial 
algebras $int(2) = u(1) $ and $int(4) =
sp(1)$ reflects the fact that the spin representations of 
$so(2)$ and $so(4)$ are complex and quaternionic (or pseudo-real)
respectively.
In our approach, using real vector spaces,
the generators of $int(n)$ are real, antisymmetric 
matrices which square to $-1$ when suitably normalised,
allowing them to be interpreted as complex structures.
\vfill \eject

\noindent
{\bf 6. The Magic Square }
\vskip 20pt

\noindent
We are now ready to return to the generic constructions of 
section 2 and use our understanding of trialities to 
convert them into something a little more magical. 
The additional ingredients we need can actually be 
listed very concisely and so we will repeat the key equations here, 
for convenience. Starting with the definitions 
of the $so(n)$ and $int(n)$ algebras, the 
proposition in section 4 states 
$$\eqalign{
tri(n) \ & = \ so(n)_\v \oplus int(n)_\v 
\cr
& = \ so(n)_\sp \oplus int(n)_\sp 
\cr
& = \ so(n)_\sm \oplus int(n)_\sm \ 
\cr
} 
\eqno (6.1)
$$
The corollary which follows tells us how to 
project onto each $so(n)$ algebra using the 
representation matrices
$$
T_{ab \vpb} = (M_A)_{ab \vpb} T_A \ , \qquad 
T_{\a \b \vpb} = (M_A)_{\a \b \vpb} T_A \ , \qquad 
T_{\ad \bd} = (M_A)_{\ad \bd} T_A \ 
\eqno (6.2) $$
We will also continue to refer to the triality 
permutation symmetry which has pervaded our work 
so far and which is evident in the results above,
but {\it without} needing to appeal directly to the 
Permutation Theorem of section 4.
\vskip 25pt

\noindent
{\bf 6.1 The first row or column}
\vskip 10pt

\noindent
To begin, we will fill in the details 
for the example discussed in the introduction, 
adopting this as a prototype.
The aim is to show that a Lie algebra 
$$ 
F_4 = so(8) \oplus \V \oplus \S_+ \oplus \S_-
\eqno (6.3) 
$$
can be obtained by following the approach of subsection 2.1 
with 
$$
\g = \h \oplus \p \ , \qquad 
\h = so(8)_\v = so(8)_\sp = so(8)_\sm \ , \qquad \p = 
\V \oplus \S_+ \oplus \S_- 
\eqno (6.4) $$
Such a reducible representation of $\h$ does not define a symmetric pair, 
so we need, firstly, to decide how to 
complete the definition of the bracket on $\p$ and, 
secondly, to prove that it satisfies the Jacobi identity.
But the choice of representation does suggest
that we should consider the following as subalgebras of $\g$, 
$$
so(9)_\v = so(8)_\v \oplus \V \ , \qquad 
so(9)_\sp = so(8)_\sp \oplus \S_+ \ , \qquad 
so(9)_\sm = so(8)_\sm \oplus \S_-  \quad
\eqno (6.5) $$
by comparison with (2.16). 

Producing $so(n{+}1) \supset so(n)$ by adding 
in the fundamental representation is a generic result in the sense that 
such a symmetric pair exists for all $n$, as discussed in detail in 
section 2. Here we are taking advantage of the fact that 
$\V$ and $\S_\pm$ can each play the role of the fundamental
representation, depending on how we choose to think of $tri(8)$.
So, while there is nothing remarkable about the existence of 
any one of these three $so(9)$ algebras in isolation,
collectively they have a very special character,
being related by triality permutations.
This symmetry ensures the existence of all three 
$so(9)$ extensions given any one of them, 
e.g.~the one corresponding to $\V$.

To specify Lie brackets, 
let $\{T_A \}$ be an orthonormal basis for $tri(8)$ and 
$\{X_a \}$,  $\{X_\a \}$,  $\{ X_{\ad} \}$
be orthonormal bases for $\V$, $\S_+$, $\S_-$ respectively
(in keeping with the notation of sections 3 to 5). 
The $so(9)$ extensions above are given by  
$$\eqalign{
[ \, T_A , \, X_a \, ] \, = \, - (M_A)_{ab} \ndg X_b 
\ , \ \qquad \qquad & 
\, [ \, X_a \ndg , \, X_b \, ] \, = \, - (M_A)_{ab} \ndg T_A \ 
\cr
[ \, T_A , \, X_\a \, ] \, = \, - (M_A)_{\a \b} \ndg X_\b 
\ , \qquad \qquad & 
[ \, X_\a , \, X_\b \, ] \, = \, - (M_A)_{\a\b} \ndg T_A \ 
\cr 
[ \, T_A , \, X_\ad \, ] \, = \, - (M_A)_{\ad \bd} \ndg X_\bd 
\ , \qquad \qquad & 
[ \, X_\ad , \, X_\bd \, ] \, = \, - (M_A)_{\ad \bd} \ndg T_A \ 
\cr 
}\eqno(6.6) $$
as in (2.4) and (2.12), but with the matrices $M$ 
now the familiar representations of 
$tri(8)$ on $\V$ and $\S_\pm$.
With $so(9)_\v$ and $so(9)\spm$ as subalgebras
of $\g$, it is sufficient to consider elements drawn 
from different subspaces $\V$, $\S_+$, $\S_-$ in order 
to complete the definition of the Lie bracket.
But since we also require invariance under $tri(8)$, 
the triality map itself allows one to write down the 
following natural expressions for the brackets
$$\eqalign{
[ \, X_a , \, X_\a \, ] & \, = \, \kappa \, 
\gamma_{a \ndg \a \ndg \ad} \ndg X_{\ad}
\cr
[ \, X_\a , \, X_\ad \, ] & \, = \, \kappa \, 
\gamma_{a \ndg \a \ndg \ad} \ndg X_{a}
\cr
[ \, X_\ad , \, X_a \, ] & \, = \, \kappa \, 
\gamma_{a \ndg \a \ndg \ad} \ndg X_{\a}
\cr
}
\eqno(6.7)$$
with an overall constant, $\kappa$, yet to be determined. 

In fact, we can arrive at these expressions far more systematically, 
by referring back to the approach of section 2.
We saw there that the general way to complete the definition of $\g$ 
as a Lie algebra is to choose an antisymmetric, trilinear, $\h$-invariant 
map $c$ on $\p$, as in (2.9). If we require that the brackets of the 
$so(9)$ algebras above are unmodified, then $c$ must give zero when 
acting on elements belonging to any two of the subspaces 
$\V$, $\S_+$, $\S_-$. But this means that the antisymmetric 
map on $\p {\times} \p {\times} \p$ is determined 
by a multilinear map on $\V{\times} \S_+ {\times} \S_-$,
and this is exactly the triality, up to the overall constant.
By comparison with (2.9) we have 
$$
( \, X_a , \, [ X_\a , X_\ad ] \, ) \, = \, 
( \, X_\a , \, [ X_\ad , X_a ] \, ) \, = \, 
( \, X_\ad , \, [ X_a , X_\a ] \, ) \, = \, 
\kappa \, \gamma_{a \ndg \a \ndg \ad} 
\eqno (6.8) $$
and this reproduces (6.7).

Do the brackets (6.6) and (6.7) satisfy the Jacobi identity? 
The remarks in section 2 ensure that this works 
automatically except for generators of type $XXX$, all belonging
to $\p$. But the existence of the $so(9)$ subalgebras also means 
that Jacobi is guaranteed whenever the generators are 
taken from the same subspace of $\p$. It is therefore 
sufficient to carry out checks for the combinations
$X_a X_\a X_\ad$, with each generator drawn from a different
subspace, and for $X_a X_b X_\a$ or similar combinations involving 
two of the three subspaces $\V$, $\S_+$, $\S_-$.

For the first combination, we need only write down the 
definitions to see that the Jacobi condition 
reduces to invariance of $\gamma$:
$$\eqalign{
&
[ \, X_a , \, [ \, X_\a , \, X_\ad \, ] \, ]
\, + \, 
[ \, X_\a , \, [ \, X_\ad , \, X_a \, ] \, ]
\, + \, 
[ \, X_a , \, [ \, X_\a , \, X_\ad \, ] \, ]  
\cr
& \qquad \ = \ - \, \kappa \, \{ \ 
(M_A)_{ab \vpb} \ndg \gamma_{\ndg b \ndg \a \ndg \ad \vpb} \, + \,  
(M_A)_{\a\b \vpb} \ndg \gamma_{a \ndg \b \ndg \ad \vpb} \, + \,
(M_A)_{\ad \bd} \ndg \gamma_{a \ndg \a \ndg \bd}
\ \} \, T_A \ = \ 0 \cr
}
\eqno (6.9) $$
For the second combination to work, we need the expressions 
$$\eqalign{
& \qquad \qquad \qquad [ \, [ \, X_a , \, X_b \, ] , \, X_\a \, ] \ = \ 
\, (M_A)_{a b} (M_A)_{\a \b} \, X_\b 
\cr
& {\rm and} \qquad 
[ \, X_a , \, [ \, X_b , \, X_\a \, ] \, ] 
\, - \, 
[ \, X_b , \, [ \, X_a , \, X_\a \, ] \, ] 
\ = \
2 \kappa^2 \, (\gamma_{ab})_{\a \b} X_\b 
\cr } 
\eqno(6.10)
$$
to match, which is equivalent to 
$$
(M_A)_{a b \vpb} (M_A)_{\a \b \vpb} \, = \, 2 \kappa^2 \ndg 
( \gamma_{ab} )_{\a \b \vpb} 
\eqno(6.11) $$
One way to confirm this is to use the identification 
$tri(8) = so(8)_\v = \wedgesq \V$ to re-label the sum on the left, 
replacing $A$ with an antisymmetric pair of vector indices $cd$ 
(along with the standard factor of $\half$); 
the definitions in (4.8a) then reveal that the 
expressions match for $\kappa = \pm 1/2$.
More satisfactorily, we can reach the same 
conclusion using (6.2): simply by 
taking matrix representations of the generators $T$ 
on each side of these relations we obtain\footnote{${}^{(3)}$}{
{\abs The same idea was used to arrive at (4.18) but using the 
fundamental representation in each case.}
}
$$\eqalign{
(M_A)_{a b \vpb} \, (M_A)_{\a \b \vpb} 
& \ = \ \half ( \gamma_{ab} )_{\a \b \vpb}
\cr
(M_A)_{\a \b \vpb} \, (M_A)_{\ad \bd} 
& \mdg \ = \ \half ( \gamma_{\a \b} )_{\ad \bd}
\cr
(M_A)_{\ad \bd} \, (M_A)_{a b \vpb} 
& \ = \ \half  ( \gamma_{\ad \bd} )_{a b \vpb}
\cr
}
\eqno(6.12)$$
These identities imply (6.11) and all 
other, similar conditions obtained by 
permuting the spaces $\V$, $\S_\pm$, provided $\kappa = \pm 1/2$,
and then the proof of the Jacobi identity is complete.

Having understood our prototype, we can proceed to give an 
analogous construction based on any one of the 
four trialities by defining Lie algebras
$$
\g = \h \oplus \p \ , \qquad 
\h = tri (n) \ , \qquad \p = \V \oplus \S_+ \oplus \S_-
\eqno(6.13) $$
with $n= 1, 2, 4$ or $8$. From the general structure (6.1) 
we now have the novel possibility of non-trivial $int(n)$ factors.
However, we can still define 
$$
so(n{+}1)_\v = so(n)_\v \oplus \V \ , \qquad 
so(n{+}1)_\sp = so(n)_\sp \oplus \S_+ \ , \qquad 
so(n{+}1)_\sm = so(n)_\sm \oplus \S_- \  \qquad 
\eqno (6.14) $$
related by triality permutations, and whose 
Lie brackets are again given by (6.6) with $\{ T_A \}$ a basis for $tri(n)$.
Because of the direct sums 
in (6.1) we obtain three inequivalent extensions 
of $tri(n)$ within $\g$, which we can summarise by the following
pattern of inclusions
$$
\matrix{ 
& & 
so(n{+}1)_\sp \oplus int (n)_\sp 
& & \cr
& \swarrow & & \nwarrow & \cr
\g & \leftarrow & 
so(n{+}1)_\v \oplus int (n)_{\v} 
& \leftarrow & tri(n) \cr
& \nwarrow & & \swarrow & \cr
& & 
so(n{+}1)_\sm \oplus int (n)_\sm 
& & \cr
}
\eqno (6.15) $$
The three intermediate subalgebras 
specify much of the bracket on $\g$, as well 
as ensuring that large parts of the Jacobi 
identity hold. The triality map can be used to 
define the remaining brackets by (6.7), as before,
and the proof of the Jacobi identity goes through unaltered
for any $n$ provided we use (6.12) and  
set $\kappa= \pm 1/2$.
\vskip 25pt

\noindent
{\bf 6.2 The complete magic square}
\vskip 10pt

\noindent
Drawing together the various lines of development 
will now bring us to the magic square. The main idea is 
to generalise the discussion of the last subsection
to two trialities, in much the same way that $so(n{+}1) \supset so(n)$ 
can be generalised to $so(n{+}n') \supset so(n) \oplus so(n')$. 
Given a pair of trialities 
$$
\gamma \, : \, \V \mdg \times \mdg \S_+ \mdg \times \mdg \S_- 
\ \rightarrow \ \R \ , 
\qquad \qquad
\gamma' \, : \, \V' \mdg \times \mdg \S'{}_\lp \mdg \times \mdg 
\S'{}_\lm \ \rightarrow \ \R
\eqno(6.16) $$
with $n$, $n' = 1, 2, 4$ or $8$ we will define a Lie algebra 
using the method of section 2, with
$$
\g = \h \oplus \p \ , 
\qquad 
\h = tri(n) \oplus tri (n') \ , 
\qquad 
\p = 
(\V \otimes \V') \oplus 
(\S_+ \otimes \S'{}_\lp) \oplus 
(\S_- \otimes \S'{}_\lm) 
\eqno(6.17) $$
Notice that we are matching up the underlying vector 
spaces in a certain way, consistent with the notation. 
This means that we should henceforth 
restrict attention to triality permutations applied 
to $tri(n)$ and $tri(n')$ simultaneously---i.e.~to 
the diagonal subgroup of the product of the permutation 
groups for $\gamma$ and $\gamma'$.

Let $\{ \ndg T_A , T_{A'} \ndg \}$ be an orthonormal basis for 
$tri(n) \oplus tri (n')$ and $\{ X_{a \ndg a'} \}$,
$\{ X_{\a\a'} \}$, $\{ X_{\ad \ad'} \}$ be orthonormal bases 
for $\V \mdg \otimes \mdg \V'$, $\S_+ \mdg \otimes \mdg \S'{}_{\! \! +}$, 
$\S_- \mdg \otimes \mdg \S'{}_{\! \! -}$ respectively. From (6.1) we have 
$$\eqalign{
\ tri(n) \oplus tri (n') 
& \ = \ 
so(n)_\v \oplus so(n')_{\v} \oplus int (n)_{\v} \oplus int (n')_{\v} 
\cr
& \ = \ 
so(n)_\sp \oplus so(n')_\sp \oplus int (n)_\sp \oplus int (n')_\sp 
\cr
& \ = \ 
so(n)_\sm \oplus so(n')_\sm \oplus int (n)_\sm \oplus int (n')_\sm 
\cr
}
\eqno(6.18) $$
which leads us to consider the algebras
$$\eqalign{
so(n{+}n')_\v & \, = \, so(n)_\v \oplus so(n')_\v \oplus (\V \otimes \V')
\cr
so(n{+}n')_\sp & \, = \, so(n)_\sp \oplus so(n')_\sp \oplus (\S_+ \otimes
\S'{}_\lp )
\cr 
so(n{+}n')_\sm & \, = \, so(n)_\sm \oplus so(n')_\sm \oplus (\S_- \otimes
\S'{}_\lm ) 
\cr
}
\eqno (6.19) $$
Each is constructed as in section 2, with brackets
(2.32) and (2.33) but with $\{ X_{i \ndg i'} \}$ 
replaced by the appropriate set of basis elements above.
Because of the direct sums in (6.18), these algebras 
lead to three different ways to extend 
$tri(n) \oplus tri(n')$ within $\g$, and hence to embeddings 
$$
\matrix{ 
& & 
so(n{+}n')_\sp \oplus int (n)_\sp \oplus int (n')_\sp 
& & \cr
& \swarrow & & \nwarrow & \cr
\g & \leftarrow & 
so(n{+}n')_\v \oplus int (n)_{\v} \oplus int (n')_{\v} 
& \leftarrow & tri(n)\oplus tri(n') \cr
& \nwarrow & & \swarrow & \cr
& & 
so(n{+}n')_\sm \oplus int (n)_\sm \oplus int (n')_\sm 
& & \cr
}
\eqno (6.20) $$
The three intermediate algebras are related by triality permutations
for $\gamma$ and $\gamma'$ simultaneously, so we could also 
infer the existence of all three from the existence of any one of
them.

The definition of the Lie bracket on $\g$ must now be completed 
by using an antisymmetric trilinear map on $\p$
which is invariant under $tri(n) \oplus tri(n')$.
But given the three $so(n{+}n')$ subalgebras in (6.20),
the map must vanish when restricted to any two of the three 
subspaces of $\p$, since brackets within 
each subspace are already determined. 
As with our prototype, this implies that an antisymmetric 
map on $\p \times \p \times \p$ reduces to a multilinear map
on its distinct subspaces, i.e.~on
$ (\V \otimes \V') \! \times \! (\S_+ \otimes \S'{}_{\! \! +}) 
\! \times \! (\S_- \otimes \S'{}_{\! \! -})$.
When the trialities are combined as $\gamma \otimes \gamma'$ 
they have exactly the properties we need; they produce a map
(compare with (2.9))
$$
( \, X_{a a'} , \, [ X_{\a \a'} , X_{\ad \ad'} ] \, ) \, = \, 
( \, X_{\a \a'} , \, [ X_{\ad \ad'}, X_{a a'}] \, ) \, = \, 
( \, X_{\ad \ad'}, \, [ X_{a a'} , X_{\a \a'} ] \, ) \, = \, 
\half \ndg \ndg \gamma_{a \ndg \a \ndg \ad} \ndg \gamma'{}_{\! \! a' \a' \ad'} 
\eqno (6.21) $$
and the resulting brackets are  
$$\eqalign{
[ \, X_{a a'} , \, X_{\a \a'} \, ] & \, = \, 
\half \ndg \ndg \ndg \gamma_{a \ndg \a \ndg \ad} \ndg 
\gamma'{}_{\! \! a'\a' \ad'} \, X_{\ad \ad'}
\cr
[ \, X_{\a \a'} , \ndg X_{\ad \ad'} \ndg ] & \, = \, 
\half \ndg \ndg \gamma_{a \ndg \a \ndg \ad} \ndg \gamma'{}_{\! \! a' \a' \ad'} 
\, X_{aa'}
\cr
[ \, X_{\ad \ad'}, \, X_{aa'} \, ] & \, = \, 
\half \ndg \ndg \gamma_{a \ndg \a \ndg \ad} \ndg \gamma'{}_{\! \! a' \a' \ad'} 
\, X_{\a \a'}
\cr
}
\eqno(6.22)
$$ 
The only freedom is the choice of an overall constant, 
which we have fixed in advance, for simplicity.
By comparison with our earlier discussion (recovered by
taking $n'=1$) the normalisation in (6.21) and (6.22)  
corresponds to fixing the sign 
and setting $\kappa = 1/2 $ in (6.7) and (6.8).

It remains to show that all brackets on $\g$ obey the 
Jacobi condition, and the arguments are very similar to those of 
the last subsection.
Within each intermediate subalgebra in $(6.20)$, Jacobi is 
automatic, and so it is sufficient to check it for elements 
belonging to two or more different subspaces of $\p$.
Up to permutations, therefore, the relevant combinations
are 
$X_{aa'}$$X_{\a \a'}$$X_{\ad \ad'}$, and 
$X_{aa'}$$X_{bb'}$$X_{\a \a'}$.
The first possibility reduces immediately, on applying the
definitions, to invariance 
of $\gamma$ and $\gamma'$, generalising (6.9). 
For the second combination we need to compare 
$$\eqalign{
& \quad \quad [ \, [ X_{aa'} , X_{bb'} ] 
, \, X_{\a \a'} \, ] \ = \ 
(M_A)_{a b} \ndg (M_A)_{\a \b} \ndg \delta_{a' b'} \ndg X_{\b \a'} 
\ + \ (M_{A'})_{a' b'} \ndg (M_{A'})_{\a' \b'} \ndg \delta_{a b} \ndg 
X_{\a \b'} 
\cr
{\rm with} & \quad 
[ \, X_{a a'} , \, [ X_{bb'} , \, X_{\a \a'} ] \, ]  
\, - \,  
[ \, X_{bb'} , \, [ X_{aa'} , \, X_{\a \a'} ] \, ] 
\ = \  
\half (\gamma_{ab})_{\a \b} \ndg \delta_{a'b'} \ndg X_{\b \a'} 
\, + \, 
\half (\gamma'{}_{\! \! a'b'})_{\a' \b'} \ndg \delta_{ab} \ndg X_{\a \b'} 
\cr } \eqno(6.23) $$
where the last expression requires just a minor rearrangement
after applying the definitions:
$$\eqalign{
[ \, X_{bb'} , \, [ X_{a a'} , X_{\a \a'} ] \, ]  \ & = \
- \, \quarter \, 
\gamma_{a \ndg \a \ndg \ad \vpb} \, \gamma'{}_{\! \! a' \a' \ad' \vpb} \,
\gamma_{\ndg b \ndg \b \ndg \ad \vpb } \, \gamma'{}_{\! \! b' \b' \ad' \vpb } 
\, X_{\b \b' \vpb}
\cr
& = \ 
- \, \quarter \, 
( \, \delta_{ab \vpb} \delta _{\a \b \vpb} 
\, + \, (\gamma_{ab})_{\a \b \vpb} \, )
( \, \delta_{a' b' \vpb} \delta _{\a' \b' \vpb} 
\, + \, (\gamma'{}_{\! \! a'b'})_{\a' \b' \vpb} \, ) \, X_{\b \b'\vpb }
\cr
}
\eqno(6.24)
$$
using (4.10), but then we antisymmetrise in $aa'\leftrightarrow bb'$
to get the expression given. 
The quantities in (6.23) now indeed coincide, using
$$(M_A)_{a b \vpb} (M_A)_{\a \b \vpb} 
= \half ( \gamma_{ab} )_{\a \b \vpb} 
\qquad {\rm and} \qquad
(M_{A'})_{a' b' \vpb} (M_{A'})_{\a' \b' \vpb} 
= \half ( \gamma'{}_{\! \! a'b'} )_{\a' \b' \vpb} 
\eqno(6.25) $$
Thus, the Jacobi identity holds by virtue of the relations 
(6.12), which follow from (6.2), for each triality.

By taking all four possibilities for $\gamma$ and $\gamma'$ in
turn, we obtain a $4{\times}4$ square of Lie algebras. The
identification of each entry can be carried out using the results 
in section 5 and considering roots 
and weights, for instance. So with $n = n' = 8$, for example, we have 
$$
E_8 \ = \ so (8) \oplus so(8) 
\oplus (\V \otimes \V') 
\oplus (\S_+ \otimes \S'{}_\lp) 
\oplus (\S_- \otimes \S'{}_\lm) 
\eqno(6.26) $$
We will not discuss such details any further, except to note that 
the result of the construction is, in general, obviously symmetrical 
in $\gamma$ and $\gamma'$. Finally, we can record our conclusions 
as follows.
\vskip 10pt

\noindent
{\bf Magic Square Theorem}:  For any trialities 
$\gamma$ and $\gamma'$ with $n$, $n' = 1, 2, 4$ or $8$, there 
exist Lie algebras 
$$
\g (n,n') \ = \ 
tri(n) \oplus tri (n') 
\oplus (\V \otimes \V') 
\oplus (\S_+ \otimes \S'{}_\lp) 
\oplus (\S_- \otimes \S'{}_\lm) 
\eqno(6.27) $$
The results are given by the magic square (1.1), 
with $n$ and $n'$ the dimensions of the division algebras 
labelling rows and columns.
\vskip 10pt

\noindent
The nature of the relationship with division algebras will be 
described in a little more detail below.
\vskip 50pt

\noindent
{\bf 7. Summary and Further Developments}
\vskip 25pt

\noindent
{\bf 7.1 Summary}
\vskip 10pt

\noindent
The construction of the magic square given above
has the desirable feature of being entirely uniform, 
producing each entry from a single, general set of definitions. 
The Lie brackets also arise naturally from the 
work on trialities in sections 3 and 4, 
and there is very little calculation required to 
verify the Jacobi identity once the definitions 
have been applied. 
In all these respects the construction 
seems about as straightforward as one could hope for.
The end result is also very much in keeping with the 
intentions expressed in the introduction.
With $G_2$ aside, the exceptional algebras 
have been understood as emerging within some general scheme, 
yet at the same time there is something special in their nature
in that they all owe their existence to the $n=8$ triality.

We also claimed in the introduction that our approach would
give a better picture of how the `exceptional' properties of 
trialities were to be combined with more `generic' aspects 
of the construction. This distinction can be seen very clearly 
for the three embeddings of $tri(n) \oplus tri(n')$ in (6.20), which 
each correspond to a symmetric pair 
$$
so(n{+}n') \ \supset \ so(n) \oplus so(n') 
\eqno (7.1) $$
There is nothing remarkable about any 
one of these extensions individually, since such symmetric pairs 
occur for all values of $n$ and $n'$.
But the existence of three, related symmetric pairs 
is certainly something special, and depends on the underlying 
triality permutations for $n$, $n' = 1, 2, 4$ or $8$. 

The trialities are then used once more to complete the 
definition of the Lie bracket. The maps $\gamma$ and $\gamma'$ provide
a natural way to do this, up to an overall constant which is determined 
via the Jacobi condition.
This amounts to specifying the second set of embeddings in 
(6.20), which also correspond to symmetric pairs
$$
\g (n, n') \ \supset \ so(n{+}n') \oplus int(n) \oplus int(n') 
\eqno (7.2) $$
The algebra $\g(n,n')$ therefore possesses three 
$\Z_2$ gradings which are compatible, in fact, so that together 
they constitute a $\Z_2 {\times} \Z_2$ grading. 
It is easy to see this explicitly from the construction: 
if $tri(n) \oplus tri (n')$ 
and exactly one of the subspaces 
$\V \otimes \V'$ or $\S_\pm \otimes \S'{}_{\! \! \pm}$ 
are taken to have grade zero and the two remaining subspaces
are taken to have grade one, then this assignment is consistent 
with all Lie brackets, and in particular with (6.22).

The final step---checking the Jacobi identity---is simple and 
transparent (compare with the calculations required for isolated
examples in [1,8] or for the entire square in [4]). 
The reason is that we can exploit the existence of the
three copies of $so(n{+}n')$ in (6.20) to limit drastically 
the combinations of brackets which need to be considered.
It turns out that even this simple final step can be dispensed with,
in a certain sense, if we are willing to appeal to some general 
properties of spinors of $so(n{+}n')$.
We will explain this briefly, to give another perspective on 
how the magic square works.
\vskip 25pt

\noindent
{\bf 7.2 An additional perspective}
\vskip 10pt

\noindent
The spaces $\V\otimes \V'$ and $\S_\pm \otimes \S'{}_{\! \! \pm}$ 
begin life as representations of 
the algebras $so(n) \oplus so (n')$ in the magic square construction, 
so it is natural to ask how they behave with respect to 
the larger algebras $so(n{+}n')$. 
Considering $so(n)_\v \oplus so(n')_\v$, for instance, we can combine 
representations and arrange that 
$$ 
so(n{+}n')_\v \quad {\rm acts~~on} \quad \, 
(\S_+ \! \otimes \S'{}_\lp) \oplus (\S_- \! \otimes \S'{}_\lm )
\eqno (7.3) $$  
but only if the action of $\V \otimes \V'$ is defined appropriately.
This is exactly what the brackets in (6.22) achieve, 
and it fixes their normalisation up to a sign.
The Jacobi condition as discussed in (6.23) is then 
an expression of the fact that the representation 
has been extended nicely from the smaller to the larger algebra.

Once again, this behaviour is not in the least 
special for any single symmetric pair, since (7.3)
amounts to a well-known decomposition property 
of spin representations which holds in arbitrary dimensions.
We can verify this without even changing our notation,
just by relaxing some assumptions to allow general values of $n$ and $n'$.
The spin representations $\S_\pm$ and $\S'{}_{\! \! \pm}$ will then have
real dimensions $N$ and $N'$, say, but they can still be described 
in terms of maps $\gamma$ and $\gamma'$ as discussed at the end of 
section 3. For any $n$ and $n'$, the action of $so(n)_\v \oplus so(n')_\v$ 
can always be extended as in (7.3)
by using brackets 
$$
[ \, X_{aa'} , \, X_{\a \a'} \, ] \, = \, \pm \ndg \half \ndg 
\gamma_{a \ndg \a \ndg \ad} \ndg \gamma'{}_{\! \! a' \a' \ad'} 
\ndg X_{\ad \ad'} 
\ , \qquad 
[ \, X_{aa'} , \, X_{\ad \ad'} \, ] \, = \, \mp \ndg \half \ndg 
\gamma_{a \ndg \a \ndg \ad} \ndg \gamma'{}_{\! \! a' \a' \ad'} 
\ndg X_{\a \a'} 
\eqno (7.4) $$
which are thereby determined, up to an overall sign, just as in
(6.22).

The really special feature of the brackets (6.22) is that they 
also ensure 
$$ 
so(n{+}n')_\spm \quad {\rm acts~~on} \, \quad 
(\V \otimes \V') \oplus (\S_\mp \otimes \S'{}_{\! \! \mp})
\eqno (7.5) $$  
in addition to (7.3), which requires 
$n$, $n' = 1, 2, 4$ or 8. 
The existence of all three representations is guaranteed 
by the existence of any one of them using the fact that 
the map (6.21) is totally 
antisymmetric under triality permutations 
(in the sense of 
merely interchanging the roles of $\V$, $\S_+$, $\S_-$).
But if we follow the arguments of section 2, and keep in mind the 
$\Z_2 \times \Z_2$ grading mentioned above, we find that the 
relations amongst the brackets required for all 
three representations exhaust the Jacobi condition.
Previously, we were able to dispense with much of the Jacobi property 
simply by noting the existence of three $so(n{+}n')$ algebras.
Now we have learnt that what remains is nothing more nor
less than the realisation of the representations (7.3) and
(7.5) via Lie brackets within $\g (n,n')$.

This brings us back, full circle, to the discussion in the
introduction about alternative approaches to our prototype, $F_4$.
There we noted that $so(9)_\v = so(8) \oplus \V$ had a representation 
on $\S = \S_+ \oplus \S_-$ but a further (un-illuminating) calculation 
was required to show directly that we ended up with a Lie algebra.
Now that we have understood that these Lie brackets originate from (6.8),
triality permutations imply that they must 
also give rise to representations of $so(9)_\spm$ on $\V \oplus
\S_\mp$. The Jacobi identity then follows in its entirety.

With hindsight, we can also see 
how the following specialisation of the approach 
in section 2 provides a template 
for the magic square. Consider building a Lie algebra 
$$ \g = \h \oplus \p \qquad {\rm with} \qquad
\p = \p_1 \oplus \p_2 \oplus \p_3
\eqno (7.6) $$
and a representation of $\h$ given on each subspace $\p_k$.
Suppose these representations define subalgebras  
$$
\g_1 = \h \oplus \p_1 \ , \qquad 
\g_2 = \h \oplus \p_2 \ , \qquad
\g_3 = \h \oplus \p_3 
\eqno(7.7) $$
such that $\g_k \supset \h$ is a symmetric pair for each $k$.
This restricts the $\h$-invariant map 
$\wedge^{\mdg 3 \ndg} \p \rightarrow \R$ used to complete 
the definition of the Lie bracket: it must give zero when applied 
to two or more elements drawn form the same subspace $\p_k$ and it
is therefore obtained by antisymmetrising some 
$\h$-invariant multilinear map 
$\p_1 \times \p_2 \times \p_3 \rightarrow \R$. 
But this in turn implies that 
the resulting bracket gives $\g$ a $\Z_2 \times \Z_2$ grading,
with each $\g \supset \g_k$ also being a symmetric pair. 
Moreover, the arguments of section 2 show that 
once the representations and multilinear map have been specified, 
the Jacobi property holds iff 
(i) the symmetric pairs $\g_k \supset \h$ exist; and (ii) 
the representations of $\h$ extend to representations of $\g_k$ 
via the bracket, so that
$$ \g_1  \ \ \ {\rm acts~on} \ \ \ \p_2 \oplus \p_3 \ , \qquad 
\g_2  \quad {\rm acts~on} \quad \p_3 \oplus \p_1 \ , \qquad 
\g_3  \quad {\rm acts~on} \quad \p_1 \oplus \p_2 
\eqno(7.8) $$ 

From this standpoint we see, once again, how the 
triality permutation symmetry of the multilinear map 
$\gamma \otimes \gamma'$ underlies the construction of 
$\g(n,n')$. It is this symmetry which 
ensures that (6.21) (with the normalisation chosen correctly) leads to 
three interlocking representations (7.3) and (7.5)
and the Jacobi identity then holds automatically, from the arguments above.
Furthermore, the existence of all three representations follows from 
the existence of any one of them.
This is another instance of how a `generic' property, 
in this case the decomposition or extension of 
spin representations, can be combined with triality permutations 
when $n$, $n' = 1, 2, 4$ or $8$ to provide a simple understanding 
of a rather non-trivial result.
\vskip 25pt

\noindent
{\bf 7.3 Division algebras and $G_2$}
\vskip 10pt

\noindent 
Although $G_2$ is the only simple exceptional algebra which does not
appear in the magic square, it too can be understood in terms of 
trialities, and in terms of the $n=8$ triality in particular. 
This turns out to be closely related to how we make the transition
from a triality to a division algebra---something we also need to 
discuss to make contact with the body of work in [4] and to 
justify labelling the entries of the square by division algebras in 
the first place. These matters are dealt with in [1] and [3], so our
treatment will be brief. In consulting these sources, however, it may
help to compare the new definition adopted in this paper with the
original version, which we have not needed up till now. 
A {\it normed triality} is defined in [1] to be a map $\gamma$ 
as in (3.1a) which obeys 
$ | \, \gamma( u_\v , u_\sp , u_\sm ) \, | \leq | u_\v || u_\sp || u_\sm |$
and such that, if any two vectors are given, there is 
a non-zero choice of the third vector for which the
bound is attained. It is a simple matter to show that this is
equivalent to our definition of a triality,
and with this understood there should be no difficulty 
in filling in the details in the following summary.

Given a triality $\gamma$ and any element $e_\v$
of unit norm in $\V$, the subgroup of $tri(n)$ which fixes this element is 
$so(n{-}1)_\v \oplus int (n)_\v \subset so(n)_\v \oplus int (n)_\v =
tri(n)$. Since the map $\sigma(e_\v)$ is invertible, the choice of 
$e_\v$ also gives us a way of identifying $\S_+$ and $\S_-$.
Now pick an additional element of unit norm $e_\sm$ in $\S_-$.
The subalgebra of $tri(n)$ which fixes 
both $e_\v$ and $e_\sm$ also fixes a unit vector $e_\sp$ in $\S_+$, where 
$$\eqalign{
&e_\sp = \sigma (e_\v) e_\sm  \ \iff \
e_\v = \sigma (e_\sm) e_\sp \ \iff \ 
e_\sm = \sigma (e_\sp) e_\v \cr
& \qquad \qquad \qquad \iff \ 
\gamma( e_\v , e_\sp , e_\sm ) \ = \ 1 
\cr
}\eqno(7.9) $$
Any two of these elements determines the third,
and the subalgebra which fixes two of them therefore fixes all
three. For $n=1$ or 2, the corresponding Lie algebra 
is trivial, but in the remaining cases we obtain 
$$
su(2) \, \subset \, tri(4) 
\qquad {\rm and} \qquad 
G_2 \, \subset \, tri(8)
\eqno(7.10) $$

The choice of corresponding elements 
$e_\v$, $e_{\spm}$ allows all three vector spaces to be 
identified, by making use of the maps 
$\r (e_\v)$, $\r (e_\spm)$. It also gives a way 
of turning the result into a division algebra
$$
\V 
\, \leftrightarrow \, 
\S_+ 
\, \leftrightarrow \, 
\S_- 
\, \leftrightarrow \, 
\K
\eqno (7.11) $$
with the distinguished, identified elements, now denoted simply by
$e$, playing the role of the identity. 
Using the same symbol for identified elements in any space,
the multiplication on $\K$ can be chosen to obey 
$$
\gamma( xy , e , z) = \gamma ( x , y , z ) 
\eqno (7.12) $$
so that it is defined in terms of $\gamma$.
There is some freedom here, depending on the details of how spaces are 
identified (up to conjugation in each 
copy of the division algebra, for instance) but the multiplication can
be shown to obey the normed division algebra axioms [1].
With appropriate choices, the original 
triality map then takes the form 
$$
\gamma ( x, y, z ) = {\rm Re} ( \ndg (xy)z \ndg )
\eqno (7.13)$$
(where the real part of an element in $\K$ is given 
by its inner-product with $e$). 
Conversely, if we are given a division algebra then 
the map above can be shown to define a triality. 

In this manner, we obtain an equivalence between 
trialities with $n = 1, 2, 4, 8$ and division algebras $\R, \C, \H, \O$.
The Lie algebras in (7.9) then correspond to the automorphism groups of 
the division algebras $\H$ and $\O$ (the automorphism group for 
$\C$ is $\Z_2$ and there are no automorphisms of $\R$; in either case
the Lie algebra is trivial).
Furthermore, the magic square construction (6.27) can now be 
re-expressed 
$$
\g (\K , \K') \ = \ tri ( \K ) \oplus tri ( \K') \oplus 3 (\K \otimes \K')
\eqno(7.14) $$
and it was studied in detail in this form by Barton and Sudbery. 
For the smaller square in which $\K$ and $\K'$ are both
associative, the entire algebra $\g (\K , \K')$ can be identified with
the anti-hermitian $3 \times 3$ 
matrices over $\K \otimes \K'$ but for the octonionic 
entries in the square, the interpretation using division algebras
requires much more care and ingenuity. We refer to
[4] for full details.

This paper has been devoted to trialities 
rather than division algebras, but it is not our intention to advocate
the use of one, rather than the other, irrespective of the
circumstances. 
It can certainly be very attractive to see results expressed 
in a compact way using division algebra multiplication. 
In some respects, however, this structure can be almost too special: 
it is versatile enough to play many different roles, and so 
translating back into more conventional language is not always easy.
Dealing with division algebras may also involve 
relinquishing some manifest symmetry (through the choice of the identity
elements) as we have seen. Ultimately the two approaches are
equivalent so one can choose whichever offers greater insights,
or use them both to get complementary ways of understanding
the same things.

\vfill \eject

\noindent
{\bf 7.4 Folding on the diagonal}
\vskip 10pt

\noindent
The magic square arises by  
combining the special properties of triality 
algebras $ tri(n) = so(n) \oplus int (n) $
for $n=1, 2, 4, 8$ with 
symmetric pairs $so(n{+}1) \supset so(n)$ or 
$so(n{+}n') \supset so(n) \oplus so (n')$.
It is natural to ask whether similar ideas can be 
applied to other symmetric pairs,
but if we want orthogonal subalgebras, to allow a link to 
trialities, then the only other possibilities 
are [2] 
$$
su(n) \, \supset \, so(n)
\eqno (7.15) $$
The choice of representation required here is the 
traceless symmetric tensor, so we could hope to give a direct
construction 
$$
\tilde \g(n) \, = \, 
tri(n) \oplus \tilde \V \oplus \tilde \S_+ \oplus \tilde \S_-
\eqno (7.16) $$
for $n = 1, 2, 4$ or 8, 
where $\tilde \V$ and $\tilde \S_\pm$ denote the traceless, symmetric 
tensor squares of $\V$ and $\S_\pm$.
Following this route, we could use (6.1) 
and appeal to the existence of three intermediate algebras 
$$su(n)_\v = so(n)_\v \oplus \tilde \V \ , \qquad 
su(n)_\sp = so(n)_\sp\oplus \tilde \S_+ \ , \qquad 
su(n)_\sm = so(n)_\sm \oplus \tilde \S_- 
\eqno (7.17) $$
in discussing the bracket and its Jacobi property.
By comparison with (6.15) or (6.20), such an approach 
could be summarised
$$
\matrix{ 
& & 
su(n)_\sp \oplus int (n)_\sp 
& & \cr
& \swarrow & & \nwarrow & \cr
\tilde \g (n) & \leftarrow & 
su(n)_\v \oplus int (n)_{\v} 
& \leftarrow & tri(n) \cr
& \nwarrow & & \swarrow & \cr
& & 
su(n)_\sm \oplus int (n)_\sm 
& & \cr
}
\eqno (7.18) $$

The algebras $\tilde \g (n)$ can indeed be constructed in this way,
but they can also be obtained with minimal extra work from the 
results we have found already.
Consider any $\g (n,n)$ on the diagonal of the magic square,
and the map $\tau$ which {\it folds\/} it by exchanging 
the $tri(n)$ factors and the representations 
$\V \leftrightarrow \V'$ and $\S_\pm \leftrightarrow \S'_\pm$.
It is not difficult to see that $\tau$ is an automorphism 
of $\g(n,n)$, of order 2, and so its fixed-point set is 
a subalgebra. It consists of the diagonal 
$tri (n) \subset tri (n) \oplus tri (n)$ in addition to the 
symmetrised tensor squares of $\V$ and $\S_\pm$, since 
the primed and un-primed spaces are identified as 
representations of the diagonal $tri(n)$. 
We have almost reached (7.16) 
by folding $\g(n,n)$ in this way, except that the tensor 
representations are not yet traceless. To understand this last step we
will make use of the explicit forms of the brackets for $\g(n,n')$.

On identifying $\V \leftrightarrow \V'$ and 
$\S_\pm \leftrightarrow \S'_\pm$ there are bases for the tensor 
product spaces 
$\{ X_{ab} \}$, $\{ X_{\a \b} \}$,
$\{ X_{\ad \bd} \}$ and we have standard decompositions into 
trace-free and pure-trace parts:
$$\eqalign{
\half (\ndg X_{ab} \ndg + \ndg X_{ba}) \mdg \, 
= \ \mdg \mdg \mdg \tilde X_{ab} 
\ndg + \ndg {\textstyle{{1 \over n}}} \ndg \delta_{ab} \ndg \ndg X_{cc} \ \qquad 
& {\rm where} \, \qquad \tilde X_{ab} 
= \tilde X_{ba} \quad \, {\rm and} \quad \,  
\tilde X_{cc} = 0 
\cr
\half (X_{\a\b} + X_{\b\a})  = \tilde X_{\a\b} 
+ {\textstyle{{1 \over n}}} \ndg \delta_{\a\b} \ndg X_{\c\c} \qquad 
& {\rm where} \qquad \tilde X_{\a\b} 
= \tilde X_{\b\a} \quad {\rm and} \quad 
\tilde X_{\c\c} \! = 0 
\cr
\half (X_{\ad\bd} + X_{\bd\ad})  = \tilde X_{\ad\bd} 
+ {\textstyle{{1 \over n}}} \ndg \delta_{\ad\bd} \ndg X_{\cd\cd} \qquad 
& {\rm where} \qquad \tilde X_{\ad\bd} 
= \tilde X_{\bd\ad} \quad {\rm and} \quad 
\tilde X_{\cd\cd} \! = 0 
\cr
}
\eqno(7.19) $$
The sets of traceless combinations 
$\{ \tilde X_{ab} \}$, $\{ \tilde X_{\a \b} \}$,
$\{ \tilde X_{\ad \bd} \}$ provide bases for 
$\tilde \V$, $\tilde \S_+$, $\tilde \S_-$ respectively.
The pure trace combinations $\{ \ndg X_{aa}, \ndg X_{\a \a}, \ndg
X_{\ad \ad} \ndg \} $
constitute an $so(3)$ subalgebra, 
since the brackets (6.22) imply  
$$\eqalign{
[ \, X_{a a} , \, X_{\a \a} \, ] & \, = \, 
\half \ndg n \ndg  X_{\ad \ad}
\cr
[ \ndg X_{\a \a} , \mdg \, X_{\ad \ad} \, ] & \, = \, 
\half  \ndg n \ndg 
X_{aa}
\cr
[ \, X_{\ad \ad}, \, X_{aa} \, ] & \, = \, 
\half \ndg n \ndg X_{\a \a}
\cr
}
\eqno(7.20) 
$$
But, in addition, this subalgebra actually commutes with everything 
in $tri(n)$ or in $\tilde \V$, $\tilde \S_\pm$; for 
example
$$\eqalign{
[ \ndg \ndg X_{c c} , \ndg \tilde X_{\a \b} \ndg ] \ & = \ 
\half \ndg  \gamma_{c \ndg \a \ndg \ad \vpb} \ndg \gamma_{c \ndg \b
\ndg \bd} \, \tilde X_{\ad \bd} 
\ = \ \half \ndg \delta_{\a \b} \ndg \tilde X_{\ad \ad} \ = \ 0 
\cr
[ \, X_{c c} , \, \tilde X_{a b} \, ] \ & = \, - \ndg 
(M_A)_{ca} \ndg \delta_{cb} \ndg T_A \, - \, (M_A)_{cb}
\ndg \delta_{ca} \ndg T_A \ = \ 0 
\cr
}
\eqno(7.21)
$$ 
We therefore reach the following conclusion.
\vskip 6pt

\noindent 
{\bf Proposition:} The subalgebra of $\g (n,n)$ which is 
fixed by the folding automorphism is a direct sum:
$$
\g(n,n) \ \supset \ \tilde \g(n) \oplus so(3) 
\eqno (7.22) $$
with $\tilde \g (n)$ given by (7.16) and (7.18).
\vskip 6pt

\noindent 
For $n = 1$ this is all rather trivial, but for
the other three trialities we obtain interesting
results. Labelling the cases by the corresponding division algebra,
in the traditional way, we have 
$$
\matrix{
\K && \C  & \H & \O \cr
\tilde \g (\K) & \ & \ \ su(3) \ \ & \ \ so(9) \ \ & \ \ E_7 \ \ \cr
}    
\eqno (7.23) $$
In particular, this gives a nice triality description of 
$$
E_7 = so(8) \oplus \tilde \V \oplus \tilde \S_+ \oplus \tilde \S_-
\eqno (7.24) $$
\vskip 50pt

\noindent
{\centerline {\bf Acknowledgements } 
\vskip 15pt
\noindent
I am indebted to Alastair King for
drawing my attention to [4] and for a number of extremely helpful 
conversations as this work progressed. I am also very grateful 
to Tony Sudbery for his perceptive remarks on both the content 
and presentation of the paper. 
This research is supported in part by Gonville and Caius
College, Cambridge.

\vfill \eject

\noindent
\centerline{\bf Appendix}
\vskip 25pt
\noindent
We have already shown in section 3 that trialities exist only if 
$n = 1, 2, 4$ or $8$. Here we elaborate on how 
existence and uniqueness in these cases is related to 
standard results in the theory of spin and Clifford algebra representations,
deriving what we need from first principles.

Any triality $\gamma$ can be related to real matrices $\r_a$ by 
$$
(\r_a)_{\a \ad} \, = \, \gamma_{a \ndg \a \ndg \ad}
\eqno ({\rm A}.1) $$
with the conditions (3.6a) and (3.6b) equivalent to 
$$
\r^{\phantom{1}}_a \r_b^{\ndg \T} \ndg  + \ndg 
\r^{\phantom{1}}_b \r_a^{\ndg \T} \, = \, 2 \ndg \delta_{ab}
\qquad \qquad 
a, b = 1, 2, \ldots , n
\eqno({\rm A}.2) $$
Conversely, any set of $n{\times}n$ real matrices which 
obey this define a triality via (A.1).
It is convenient to consider the real symmetric matrices
$$
\Gamma_a = \pmatrix{0 & \r_a \cr \r_a^{\ndg \T} & 0 \cr }
\qquad {\rm on} \qquad \S = \S_+ \oplus \S_-
\eqno({\rm A}.3) $$
so that (A.2) is holds iff 
$$
\{ \ndg \Gamma_a , \, \Gamma_b \ndg \} \, = \, 2 \ndg \delta_{ab}
\qquad \qquad 
a, b = 1, 2, \ldots , n
\eqno({\rm A}.4) $$
We will also consider more general, hermitian matrices 
$$
\Gamma_a = \pmatrix{0 & \Sigma_a \cr \Sigma_a^{\, \dagger} & 0 \cr }
\qquad {\rm on} \qquad \Omega = \Omega_+ \oplus \Omega_-
\eqno({\rm A}.5) $$
where $\Omega_\pm$ are complex vector spaces, 
and then (A.4) is equivalent to 
$$
\Sigma^{\phantom{\dagger}}_a \ndg \Sigma_b^{\, \dagger}
\, + \, 
\Sigma^{\phantom{\dagger}}_b \ndg \Sigma_a^{\, \dagger} 
\, = \, 2 \delta_{ab}
\qquad \qquad 
a, b = 1, 2, \ldots , n
\eqno({\rm A}.6) $$

There are two distinct ways to relate 
(A.6) to (A.2). It may be simply that the matrices 
$\Sigma_a$ can be chosen to be real, enabling them to be 
identified with $\sigma_a$ (in some basis) 
with $\Omega_\pm$ the complexifications of $\S_\pm$.
If this is not possible, then we 
can still resort to re-writing each complex $\Sigma_a$ as a larger 
real matrix $\sigma_a$ by 
choosing to regard $\Omega_\pm$ as real vector spaces $\S_\pm$ of twice 
the dimension. We can also carry this out in reverse provided 
there exist suitable complex structures on $\S_\pm$ 
which are respected by the maps $\sigma_a$, so ensuring
that they can be re-expressed as complex linear maps on $\Omega_\pm$.
From our discussion of $int(n)_\v$ in section 4, we know that its
generators can be regarded as pairs of antisymmetric matrices 
$\lambda_\spm$ acting on $\S_\pm$, and obeying
$$
\lambda_\sp \ndg \sigma_a \, = \, \sigma_a \ndg \lambda_\sm
\eqno ({\rm A}.7)$$
(in (4.15) $\lambda_{\a \b \vpb}$ and $\lambda_{\ad \bd}$ are
the components of $\lambda_\spm$ ). 
From section 5, we also know that these matrices can
be normalised so that $\lambda_\spm^2 = - 1$,
and we therefore have exactly the kinds of complex structures just
discussed whenever $int(n)$ is non-trivial, i.e.~for $n = 2$ or $4$.
\vskip 10pt

\noindent
{\bf Proposition:} (a) There exist $2N{\times}2N$ complex hermitian matrices 
$\Gamma_a$ which obey (A.4) with $N = 2^{\ell-1}$ for  
any even integer $n = 2 \ell$. Any two sets of 
such matrices are related by a transformation
$$
\Gamma_a \ \mapsto \ \U \ndg \Gamma_a \U^{-1} \qquad {\rm where} \qquad
\U^{\ndg \dagger} = \U^{-1}
\eqno({\rm A}.8) $$
\hfill \break
(b) The matrices in part (a) have a block form (A.5) with 
$\Omega_\pm$ eigenspaces of 
$$
\hat \Gamma \, \equiv \,  
\Gamma_{1} \Gamma_2 \ldots \Gamma_{2 \ell}  
\, = \, \xi \pmatrix{1 & 0 \cr 0 & -1 \cr}
\qquad {\rm where} \qquad 
\xi \ndg = \pm \ndg i^{\ndg \ndg \ell}
\eqno({\rm A}.9) $$
Any two sets of such matrices with the same sign for $\xi$ are 
related by (A.8) with $\U$ block diagonal, or 
$$
\Sigma_a \ \mapsto \ U_+ \ndg \Sigma_a \ndg U_-{}^{\! \! \dagger} 
\qquad {\rm where} \qquad 
U_\pm{}^{\! \! \dagger} = U_\pm{}^{\! \! -1}
\eqno ({\rm A}.10) $$
(c) If, in addition, $n=2 \ell$ is a multiple of 8, then we can choose 
the matrices $\Gamma_a$ and $\Sigma_a = \sigma_a$, real.
Any two sets of such matrices with the same sign for $\xi$ are 
related by (A.10) with $U_\pm = R_\pm$ real, or
$$
\sigma_a \ \mapsto \ R_+ \ndg \sigma_a \ndg R_-{}^{\! \! \T} 
\qquad {\rm where} \qquad
R_\pm{}^{\! \! \T} = R_\pm{}^{\! \! \! -1}
\eqno ({\rm A}.11) $$
\vskip 10pt

\noindent
Proof: (a) Existence and uniqueness up to transformations (A.8) 
can be established by constructing joint eigenvectors 
for the mutually commuting anti-hermitian matrices 
$$
\Gamma_{12} , \ \Gamma_{34} , \, \ldots , \ \Gamma_{2 \ell{-}1 \, 2 \ell}
\eqno ({\rm A}.12) $$
which each have eigenvalues $\pm i$. 
This basis will define a standard complex form for matrices obeying
(A.4) and thereby also show that such matrices exist.  
First note that the combinations 
$$
A^\pm_r = \half (\Gamma_{2r-1} \pm i \Gamma_{2r}) \qquad 
{\rm obey} \qquad 
\{ A^\pm_r , A^\pm_s \} = 0
\quad {\rm and} \quad 
\{ A^+_r , A^-_s \} = \delta_{rs} \qquad r, s  = 1, 2, \ldots , \ell
\eqno ({\rm A}.13)$$
(and so behave like fermion creation and annihilation operators).
A unit vector $\Psi$ in $\Omega$ is a joint eigenvector of the matrices 
(A.13) with all eigenvalues $-i$ iff $A^-_r \Psi = 0$ (it is the
analogue of the ground state for the multi-fermion system) 
and this vector can be extended to a basis for $\Omega$ consisting of 
$$
(A^+_1)^{m_1} \! \ldots (A^+_{\ell})^{m_{\ell}} \, \Psi \qquad \qquad 
m_r = 0~{\rm or}~1
\eqno({\rm A}.14) $$ 
(each $m_r$ is like a fermion occupation number).
The vectors (A.14) are orthonormal by virtue of 
(A.13) and the fact that every $\Gamma_a$ is hermitian.
The action of each $A^\pm_r$ on the expressions in (A.14) 
determines their matrices with respect to this
basis---we note that they are all real---and 
the standard form for $\Gamma_a$ is then obtained from (A.13). 
Any two sets of matrices $\Gamma_a$ are unitarily equivalent to
this standard form and hence to one another.
\hfill \break
\noindent
(b) Consider the subspaces spanned by vectors in (A.14) 
for which $\sum_r m_r$ is even or odd;
define one to be $\Omega_+$ and the other to be $\Omega_-$, in some
order. These subspaces are mapped to one another by every $A^\pm_r$ and
hence by each $\Gamma_a$; they are also eigenspaces of 
$\hat \Gamma$, with the value of $\xi$ consistent with
$\hat \Gamma^2 = (-1)^\ell$. 
Which of the subspaces $\Omega_\pm$ corresponds to which eigenvalue of
$\hat \Gamma$ cannot be determined from the block form
for $\Gamma_a$ alone, leaving the sign ambiguity in $\xi$.
For any two sets of matrices related by (A.8) we have 
$\U \hat \Gamma \U^\dagger = \hat \Gamma$, but then if each set has
the same value of $\xi$, this implies that $\U$ must be block diagonal.
\hfill \break
\noindent
(c) If matrices $\Gamma_a$ meet the conditions in part (a) then so do
their complex conjugates. The uniqueness result therefore 
implies that there exists a unitary matrix $B$ with 
$$
\B \ndg \Gamma_a{}^{\! \! *} 
\ndg \B^{-1}  \, = \, \Gamma_a \qquad {\rm and} \qquad 
\B \ \mapsto \ \U \ndg \B \ndg \U^\T 
\eqno({\rm A}.15) $$
with the latter transformation accompanying (A.8), for consistency.
With respect to the basis (A.14), 
$\Gamma_a$ is real and symmetric if $a$ is odd, 
and imaginary and antisymmetric if $a$ is even 
(since each $A^\pm_r$ is real) and so we may  
choose $\B = \Gamma_2 \Gamma_4 \ldots \Gamma_{2 \ell}$, because 
$\ell$ is even.
This commutes with $\hat \Gamma$ and both matrices are real, 
symmetric and square to $1$ since $\ell$ is in fact a multiple of 4. 
They can therefore be simultaneously diagonalised by (A.8) and (A.15) 
with $\U$ real and orthogonal in these circumstances. 
But having made $\B$ diagonal with
entries $\pm 1$, a further transformation (A.8) and (A.15) with 
$\U$ complex and diagonal can be chosen to produce $\B =1$, 
which gives the result.
\vskip 10pt

\noindent
{\bf Corollary:} There exist trialities with $n = 2, 4$ or 8, each 
unique up to isomorphism, as in the Classification Theorem. 
\vskip 5pt

\noindent
Proof: When $n= 2 \ell = 8$, part (c) above implies there exist real
$N{\times}N$ matrices $\sigma_a$ obeying (A.2) with 
$N = 2^{\ell -1} = 8$ and so (A.1) defines a triality.
Conversely, from any two trialities with $n=8$ we can construct 
sets of matrices $\sigma_a$ as in part (c). 
An orthogonal transformation on $\V$
which exchanges two basis elements will change the sign of $\xi$, 
so by using this if necessary we can assume that the signs match
for each set. But then part (c) implies that the trialities are related by 
$$\gamma_{a \ndg \a \ndg \ad \vpb} \ \ \mapsto \ \  
R_{\a \b \vpb} \, R_{\ad \bd} \, \gamma_{a \ndg \b \ndg \bd} 
\eqno({\rm A}.16) $$
with $R_{\a \b \vpb}$ and  $R_{\ad \bd}$ the entries of the 
matrices $R_\pm$.

When $n = 2 \ell = 2$ or $4$, part (b) implies that there exist 
complex $N{\times}N$ matrices $\Sigma_a$ obeying (A.6) 
which in turn define real $2N{\times}2N$ matrices $\sigma_a$ obeying
(A.2), with $2N = 2^{\ell} = n$ for these two values.
Hence (A.1) again defines a triality in these cases.
When the dimensions of the spaces are doubled to pass from 
complex to real notation in this fashion, 
any unitary maps $U_\pm$ on $\Omega_\pm$ clearly become 
orthogonal maps $R_\pm$ on $\S_\pm$, and so we can use (A.10) 
to establish uniqueness by following the same argument as in the 
$n=8$ case. The only subtlety is that
if we are given two trialities expressed as sets of real matrices 
$\r_a$, then we must be able to re-write these as complex matrices 
$\Sigma_a$ in order to apply the uniqueness result in part (b).
But this is ensured by our remarks on the existence
of complex structures which the maps $\sigma_a$ respect,
as in (A.7), when $n= 2$ and $n = 4$.

\vfill \eject

\centerline{\bf REFERENCES}
\vskip 30pt

\item{[1]} J.F.~Adams, {\it Lectures on Exceptional Lie Groups},
editors Z.~Mahmud and M.~Mimira \hfil \break
 University of Chicago Press (1996)
\vskip 10pt

\item{[2]} S.~Helgason, {\it Differential Geometry, Lie Groups and
Symmetric Spaces}, Academic Press (1978)
\vskip 10pt

\item{[3]} J.C.~Baez, {\it The Octonions}, 
\hfill \break
Bull.~Amer.~Math.~Soc.~39(2) 145-205 (2002), 
{\tt arXiv:math/0105155 [math.RA]} 
\vskip 10pt

\noindent
\item{[4]} C.~Barton and T.A.~Sudbery, {\it Magic Squares and Matrix Models
of Lie Algebras}, \hfill \break
Adv.~Math.~180(2) 596-647 (2003), {\tt arXiv:math/0203010 [math.RA]}; 
\hfill \break 
{\it Magic Squares of Lie Algebras}, {\tt arXiv:math/0001083 [math.RA]} 
\vskip 10pt

\noindent
\item{[5]} J.M.~Evans, {\it Supersymmetric Yang-Mills Theories and Division
Algebras}, \hfill \break
Nucl.~Phys.~{\bf B298} (1988) 92-108
\vskip 10pt

\noindent
\item{[6]} T.~Kugo and P.K.~Townsend, {\it Supersymmetry and the Division
Algebras}, \hfill \break 
Nucl.~Phys.~{\bf B221} (1983) 357-380
\vskip 10pt

\noindent
\item{[7]} A. Sudbery, {\it Division Algebras, (Pseudo)Orthogonal Groups and
Spinors}, J.~Phys.~{\bf A17} (1984) 939-955; 
\hfill \break
K.-W.~Chung and A.~Sudbery, {\it Octonions and the Lorentz and
Conformal Groups of Ten-dimensional Space-Time}, Phys.~Lett.~{\bf
B198} (1987) 161-164
\vskip 10pt

\item{[8]} M.B.~Green, J.H.~Schwarz and E.~Witten, {\it Superstring Theory\/}, 
Cambridge University Press (1987) 
\vskip 10pt

\item{[9]} S.~Weinberg, {\it The Quantum Theory of Fields III: Supersymmetry},
Cambridge University Press (2000) 
\vskip 10pt

\item{[10]} P.~Deligne et al., {\it Quantum Fields and Strings: A course for
Mathematicians}, Amer.~Math.~Soc.~(1999) 
\vskip 10pt

\item{[11]} G.~Sierra, {\it An Application of the Theories of Jordan Algebras
and Freudenthal Triple Systems to Particles and Strings},
Class.~Quantum Grav.~{\bf 4} (1987) 227-236
\vskip 10pt

\noindent
\item{[12]} C.~Manogue and A.~Sudbery, {\it General Solutions of Covariant
Superstring Equations of Motion}, \hfill \break 
Phys.~Rev.~{\bf D12} (1989) 4073-4077
\vskip 10pt

\noindent
\item{[13]} A.~Ach\'ucarro, J.M.~Evans, P.K.~Townsend and D.L.~Wiltshire, {\it
Super p-branes}, \hfill \break
Phys.~Lett.~{\bf B198} (1987) 441-446
\vskip 10pt

\noindent 
\item{[14]} J.M.~Figueroa-O'Farrill, {\it 
A Geometric Construction of the Exceptional Lie Algebras $F_4$ and
$E_8$}, \hfill \break
{\tt arXiv:0706.2829 [math.DG]}
\vskip 10pt

\noindent
\item{[15]} C.~Pedder, J.~Sonner and D.~Tong, {\it The Berry Phase of
D0-Branes}, \hfill \break 
JHEP 0803:065 (2008), {\tt arXiv:0801.1813 [hep-th]}
\vskip 10pt

\noindent
[16] J.C.~Baez and J.~Huerta, {\it Division Algebras and
Supersymmetry}, {\tt arXiv:0909.0551 [hep-th]}
\vskip 10pt

\noindent
[17] I.R.~Porteous, {\it Clifford Algebras and the Classical Groups}, 
Cambridge University Press (1995) 
\vskip 10pt

\noindent
[18] A.L.~Onishchik and E.B.~Vinberg,  {\it Lie Groups and Lie
Algebras III}, Springer-Verlag (1991)

\bye